\documentclass{article}

\usepackage{PRIMEarxiv}

\usepackage[utf8]{inputenc} 
\usepackage[T1]{fontenc}    
\usepackage{hyperref}       
\usepackage{url}            
\usepackage{booktabs}       
\usepackage{amsfonts}       
\usepackage{nicefrac}       
\usepackage{microtype}      
\usepackage{lipsum}
\usepackage{fancyhdr}       
\usepackage{graphicx}       
\graphicspath{{media/}}     
\usepackage{natbib}
\usepackage{amsmath}
\usepackage{amsfonts}
\usepackage{amssymb}
\usepackage{siunitx}
\usepackage{mathrsfs}
\usepackage{hyperref}
\usepackage{cleveref}
\usepackage{tikz}
\usepackage{subcaption}
\usetikzlibrary{external}
\usepackage{stackengine}
\usepackage{scalerel}
\usepackage{cancel}
\usepackage{comment}
\usepackage{orcidlink}
\usepackage{physics}
\newcommand{\bs}[1]{\vb*{#1}} 

\newif\ifdiffmode
\ifdefined\diffmodeflag
  \diffmodetrue
\else
  \diffmodefalse
\fi
\DeclareRobustCommand{\reallywideoverarc}[1]{\ThisStyle{%
  \setbox0=\hbox{$\SavedStyle#1$}%
  \stackengine{-1.9\LMpt}{$\SavedStyle#1$}{%
    \stretchto{\scaleto{\SavedStyle\mkern.2mu\frown}{.4\wd0}}{.6\ht0}%
  }{O}{c}{F}{T}{S}%
}}

\DeclareRobustCommand{\arc}[1]{\ensuremath{\reallywideoverarc{#1}}}

\usepackage{physics}
\usepackage{scalerel}
\usepackage{stackengine}

\newcommand{\Space}[1]{\ensuremath{\mathscr{#1}}}
\DeclareMathAlphabet{\mathscrbf}{OMS}{mdugm}{b}{n}
\newcommand{\SpaceV}[1]{\ensuremath{\mathscrbf{#1}}}

\DeclareMathOperator{\interior}{int}
\DeclareMathOperator{\DifferentialOperator}{\bs{\partial}}
\DeclareMathOperator*{\assembly}{\scalerel*{\mathbb{A}}{\sum}}
\DeclareMathOperator*{\argmin}{arg\,min}

\makeatletter
\newcommand*{\rom}[1]{\textit{\expandafter\@slowromancap\romannumeral #1@}}
\makeatother

\newcommand{\SIFmodeI}{K_{\rom{1}}}
\newcommand{\SIFmodeII}{K_{\rom{2}}}
\newcommand{\SIFmodeIII}{K_{\rom{3}}}

\title{$\partial^2 ( \mathrm{TO} ) $: A Dual Topological Derivative-Based Enriched Topology Optimization for Fracture Mitigation in 3-D Brittle Solids}

\author{
  Majd Kosta, \\
  Affiliation \\
  Univ \\
  City\\
  \texttt{\{Author1, Author2\}email@email} \\
   \And
  Shangru Liu \\
  Affiliation \\
  Univ \\
  City\\
  \texttt{email@email} \\
  \And
    Alejandro M. Aragón \\
  Affiliation \\
  Univ \\
  City\\
  \texttt{email@email} \\
}

\author{
  Majd Kosta \hspace{1cm}  Shangru Liu  \hspace{1cm} Alejandro M. Aragón \\
  \\
  Faculty of Mechanical Engineering, Delft University of Technology, Mekelweg 2, 2628 CD Delft, The Netherlands \\ 
  \texttt{\{Alejandro M. Aragón\} a.m.aragon@tudelft.nl} \\
}

\author{Majd Kosta \,\orcidlink{0000-0001-5450-8860} \And
Shangru Liu \And
Alejandro M. Aragón \,\orcidlink{0000-0003-2275-6207}
\thanks{Corresponding author. E-mail address: \texttt{\href{mailto:a.m.aragon@tudelft.nl}{a.m.aragon@tudelft.nl}} (A.M.~Aragón)}
	 \AND
	 \\[-1em] 
	 Faculty of Mechanical Engineering, Delft University of Technology\\
	 Mekelweg 2, 2628 CD, Delft, Zuid-Holland,
	 The Netherlands\\	
}

\begin{document}
\maketitle

\begin{abstract}
We propose a fracture-mitigation topology optimization framework for 3-D brittle solids. The topology is described by a level set function parameterized by radial basis functions, and the structural response is computed using an interface-enriched finite element formulation. Dual topological derivatives serve two purposes. First, they are used to nucleate holes within the solid during the optimization process. Second, they are used to evaluate energy release rates (ERRs) along the entire boundary, requiring only the stress field from a single enriched finite element analysis of the uncracked geometry. For this purpose, penny-shaped cracks are assumed to nucleate using the maximum hoop stress criterion, at the locations of enriched nodes introduced along the boundary for accurate finite element analysis. Because ERR estimates depend sensitively on stress accuracy, we compute a nodal stress field using a non-local stress-recovery procedure. The topology optimization objective aggregates the boundary ERRs using a $p$-mean function. Three-dimensional numerical examples, including the commonly studied L-bracket benchmark problem, demonstrate the capability of the proposed framework.
\end{abstract}


\section{Introduction}

Fracture remains one of the most critical failure mechanisms in engineered structures. In brittle materials, where fracture may occur suddenly and with little warning, failure can have devastating consequences, motivating the design of structures that mitigate fracture risk. This is especially true as modern engineering pushes material systems closer to their performance limits and structural designs grow increasingly complex.

In such cases, design approaches based primarily on empirical intuition and trial-and-error become increasingly inadequate, leaving computational design methodologies not merely as attractive alternatives, but as essential tools for maximizing structural performance and service life. Topology optimization (TO) has become a widely used computational methodology for structural design~\citep{bendsoe2013topology,sigmund2013topology,van2013level}. In the context of failure-resistant design, TO formulations are commonly distinguished by the failure descriptor used to guide the optimization, including stress-based, damage-based, and fracture-mechanics-based criteria.

Stress-based TO constitutes one of the most established approaches for failure-resistant structural design. The central difficulty in stress-constrained TO stems from the local nature of stress constraints: enforcing admissible stress levels throughout the domain may introduce one or more local stress constraints per element, leading to a large number of constraints. This difficulty is further compounded by the nonlinear and singular behavior of stresses in low-density regions~\cite{Cheng1997relaxation, duysinx1998topology}, which has motivated the development of relaxation, aggregation, and alternative constraint-handling strategies. For instance, \citet{yang1996stress} used an aggregation strategy based on the Kreisselmeier–Steinhauser (KS) function~\citep{kreisselmeier1980systematic} to control peak von Mises stresses in density-based formulations. Such global stress measures, together with related aggregation functions such as the $p$-norm and $p$-mean~\citep{duysinx1998new,holmberg2013stress}, have been widely used to reduce the number of constraints and approximate the maximum stress. However, replacing a spatially distributed stress field with a single scalar quantity inevitably compromises local control, and localized stress violations may be smoothed out depending on the choice of aggregation parameters. To improve local stress control, \citet{wang2018heaviside} proposed an aggregation of Heaviside functions, whereas \citet{senhora2020topology} introduced an aggregation-free augmented Lagrangian formulation in which stresses are incorporated directly into the objective function. Beyond the von Mises criterion, stress-based TO has also been extended to other yield descriptions, including Drucker–Prager\citep{bruggi2012topology} and unified frameworks capable of representing multiple criteria such as Tresca, Mohr–Coulomb, Bresler–Pister, and Willam–Warnke~\citep{giraldo2020unified}. Despite these advances, stress-based criteria remain indirect proxies for brittle fracture, since they do not explicitly quantify the crack-driving force associated with potential crack nucleation and propagation.

Damage-based TO provides an alternative route to failure-resistant design by incorporating material degradation directly into the optimization problem. Early work includes \citet{bendsoe1998method}, who constrained damage using a continuum damage model for brittle materials. For reinforced concrete, \citet{amir2013reinforcement} combined a gradient-enhanced continuum damage model with truss topology optimization to determine reinforcement layouts. \citet{James:2014} proposed a framework to mitigate failure by using a quasi-static nonlocal brittle damage model; they optimized structures for minimum weight to single-load cases, and later extended that work to multiple loading conditions through a damage-superposition approach~\citep{james2015topology}. Elastoplastic damage models have also been adopted for the design of energy-absorbing and failure-resistant structures~\citep{li2017topology,alberdi2017topology,li2018failure}. Nevertheless, three-dimensional damage-based TO formulations remain comparatively scarce. Examples include 3-D reinforcement-layout optimization in concrete using gradient-enhanced strain-softening damage models~\citep{amir2013reinforcement} and fatigue-constrained topology optimization based on the Palmgren–Miner linear damage accumulation hypothesis~\citep{chen2020fatigue}. As with stress-based TO, damage-based formulations generally do not explicitly compute fracture-related quantities, but rather model failure through damage variables.

Unlike stress- or damage-based criteria, fracture mechanics accounts for the combined effects of loading, flaw size and geometry, and material fracture toughness. Fracture-based criteria are therefore better suited to assessing the influence of cracks in a structure~\citep{anderson2017fracture}. Incorporating such criteria into a topology optimization methodology provides a powerful framework to design structures that are less susceptible to fracture. Three strategies have been proposed in this context: stationary cracks at a given location~\citep{kang2017topology, hu2019fracture, singh2026designing}, cracks propagating from a predefined location~\citep{russ2019topology, li2021simp, wu2020level, wu2021path, desai2022topology}, and cracks nucleating throughout the design boundary. In what follows, we focus on the latter.

Topology optimization formulations that account for multiple potential crack configurations remain rather scarce. Diffuse crack models, such as damage- and phase-field-based formulations, can nucleate multiple cracks without prescribing their location, but the cracks emerge only in critical regions. In contrast, discrete crack models require either crack-fitted meshes or enriched finite element formulations; because these models require an analysis per cracked configuration, the computational demands increase rapidly with the number of crack configurations considered. Consequently, studies that have considered multiple potential crack nucleating sites have relied on simplifying assumptions to keep computational costs tractable. For instance, \citet{challis2008fracture} proposed a topology optimization formulation for fracture resistance, whereby cracks are allowed to nucleate anywhere along the structural boundary; energy release rates (ERRs) are calculated after a single FEA by evaluating the change of elastic potential energy between two configurations differing in crack length (the so-called \textit{virtual crack extension} technique). \citet{zhang2021on} proposed a level set-based interface-enriched topology optimization framework to tailor fracture resistance in brittle solids. While they also assumed cracks nucleating along the boundary, they used topological derivatives~\citep{silva2011energy} to evaluate ERRs using the results of a single enriched FEA of the uncracked geometry. However, their methodology required an initial design pre-seeded with holes to allow topology changes, and their nonlinear level set was only linearly represented in the analysis mesh, resulting in relatively rough boundaries. In addition, despite the success of topological derivatives in two-dimensional fracture-aware optimization, their use for designing 3-D brittle solids has yet to be explored.


This paper extends the work of \citet{zhang2021on} and presents a 3-D topology optimization formulation for enhancing the fracture resistance of brittle solids. Making the leap to 3-D required several methodological developments. While the topology is still represented by a level set function parameterized by radial basis functions, the projected level-set field is interpolated quadratically, thereby yielding smoother designs. Furthermore, a regularization procedure is used to maintain the quality of the level-set representation. The interface-enriched generalized finite element method (IGFEM)~\citep{Soghrati:2012} remains the analysis framework to accurately resolve the topology boundary without remeshing. To improve the accuracy of the recovered stress field, we use the stress improvement procedure (SIP) developed for 3-D linear elements by~\citet{sharma2018improved}. Cracks are modeled as half penny-shaped cracks and are assumed to nucleate normal to solid–void interfaces at the locations of enriched nodes. The crack orientation follows the maximum hoop-stress criterion, and ERRs are evaluated using 3-D topological derivatives~\citep{alidoost2020energy}. Similar to \citet{zhang2021on}, ERRs are evaluated from the results of a single enriched FEA of the uncracked domain, thereby avoiding the need to explicitly model and analyze a large number of cracked configurations. In contrast to previous work, topological derivatives are also used herein to nucleate holes in the solid domain. We refer to this dual use of topological derivatives, coupled with enriched FEA, as $\partial^2 ( \mathrm{TO} ) $. The proposed framework is demonstrated through three numerical examples. First, a shape optimization problem involving a cube containing an internal cavity under triaxial loading is considered to verify the formulation and investigate the resulting fracture-resistant cavity geometries. This example is then extended to include topological derivatives, enabling topology optimization through the nucleation of new cavities. Second, a series of shape optimization examples subjected to opening, shearing, and tearing loading modes is presented to examine the influence of different fracture mechanisms on the optimal cavity design. Finally, a three-dimensional L-bracket benchmark problem is studied to demonstrate the ability of the proposed method to improve fracture resistance in complex engineering structures.

\section{Formulation}
\label{ch:form}

\subsection{Elastostatics problem description}
\label{sec:elas}

\begin{figure}[h]
	\centering
	\includegraphics{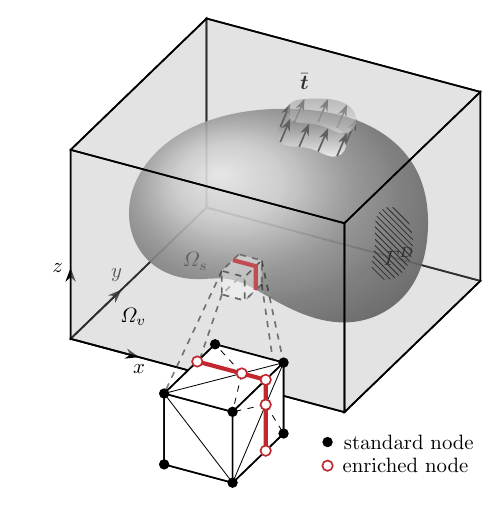}
	\caption{A domain $\varOmega$ consists of a solid $\varOmega_s$ and void $\varOmega_v$ subdomains. Dirichlet and Neumann boundary conditions are prescribed on $\varGamma^D$ and $\varGamma^N$, respectively. The inset shows a set of tetrahedral elements forming a cube, which are cut by the solid-void interface $\varGamma^I$ (shown in red). Solid lines and black dots delineate the tetrahedra. Dashed lines and red dots (enriched nodes) also delineate integration elements.}
    \label{fig:potato}
\end{figure}

Consider the open domain $\varOmega \subset \mathbb{R}^3$ shown in \Cref{fig:potato}. The domain consists of disjoint solid and void open subdomains, $\varOmega_s$ and $\varOmega_v$, respectively, such that $\varOmega_s \cap \varOmega_v = \emptyset$ and $ \overline{\varOmega} = \overline{\varOmega_s \cup \varOmega_v}$. We denote the boundary of $\varOmega$ by $\varGamma := \partial\varOmega$, and the boundary of each subdomain by $\varGamma_i := \partial\varOmega_i = \overline{\varOmega_i} \setminus \varOmega_i, \, i \in \left\{ s,v \right\}.$ Portions of the internal solid–void boundary may be subject to prescribed Dirichlet or Neumann boundary conditions. We therefore denote by $\varGamma^I \subseteq \varGamma_s \cap \varGamma_v$ the remaining internal interface, on which no boundary condition is prescribed and where the standard continuity and traction-equilibrium interface conditions are imposed. Dirichlet boundary conditions are prescribed on $\varGamma^D := \varGamma_s^D \cup \varGamma_v^D$, and Neumann boundary conditions on $\varGamma^N := \varGamma_s^N \cup \varGamma_v^N$. Thus, up to sets of measure zero, $\varGamma_i = \varGamma^I \cup \varGamma_i^D \cup \varGamma_i^N, \, i \in \left\{ s,v \right\}.$ Finally, outward normal vector fields $\bs n_s$ and $\bs n_v$ are defined on the solid and void boundaries, respectively, with $\bs n_s=-\bs n_v$ almost everywhere on $\varGamma^I$. We assume the subdomains and the interface are sufficiently regular, e.g., Lipschitz, so that traces and outward normals are well defined almost everywhere.

We denote the displacement field by $\bs{u}$ and write $\bs{u}_i := \left.\bs{u}\right|_{\varOmega_i}$ for its restriction to phase $i$, with the same convention used for other phase-wise quantities. The prescribed displacement field on $\varGamma^D$ is denoted by $\bar{\bs{u}}$, and its restriction to $\varGamma_i^D$ by $\bar{\bs{u}}_i$. Similarly, $\bar{\bs{t}}_i$ denotes the prescribed traction on $\varGamma_i^N$. Both the solid and void phases are modeled as linearly elastic materials, with Young’s modulus $E_i$ and Poisson’s ratio $\nu_i$. We assume $E_i > 0$, $-1<\nu_i<1/2$, and that $\varGamma^D$ has sufficient measure to remove rigid-body modes. Throughout the following, the void phase is treated as a weak elastic material, with $E_v \ll E_s$.

Omitting body forces for simplicity, the linear elastostatics boundary value problem is given by
\begin{alignat}{2}
  \div \bs{\sigma}_i &= \bs{0}  && \text{in } \varOmega_i, \label{eq:equilibrium}
\intertext{with boundary conditions,}
  \bs{u}_i &= \bar{\bs{u}}_i && \text{on } \varGamma^D_i, \\
  \bs{\sigma}_i \cdot \bs{n}_i &= \bar{\bs{t}}_i && \text{on } \varGamma_i^N,
\intertext{for $i \in \left\{s,v \right\}$, and interface conditions}
   \bs{u}_s &=  \bs{u}_v \quad && \text{on } \varGamma^I, \\
   \bs{\sigma}_s \cdot \bs{n}_{s} + \bs{\sigma}_v \cdot \bs{n}_{v} &= \bs{0} \quad && \text{on } \varGamma^I,
\end{alignat}
where $\div$ is the divergence operator and $\bs{\sigma}_i := \boldsymbol{\sigma}|_{\varOmega_i}$ is the Cauchy stress tensor. We adopt Hooke's law, and therefore $\bs{\sigma}_i = \mathbb{C}_i : \bs{\varepsilon}_i$, with $\bs{\varepsilon}_i = \frac{1}{2} \left( \grad \bs{u}_i + \grad \bs{u}_i^\intercal \right)$ the linearized strain tensor and $\mathbb{C}_i$ the fourth-order constitutive tensor.

For simplicity, we restrict the following derivation to homogeneous Dirichlet boundary conditions, i.e., $ \bar{\bs{u}}_i= \bs{0}$ on $\varGamma_i^D$.
We define the admissible displacement space
\begin{equation}
  \SpaceV{V}_0 \left( \varOmega \right) := \left\{ \bs{v} \in \left[ \Space{H}^1 \left( \varOmega \right) \right]^3, \, \trace \left( \bs{v} \right) = \bs{0} \text{ on } \varGamma^D \right\},
\end{equation}
where $\left[ \Space{H}^1(\varOmega) \right]^3$ is the vector-valued Sobolev space of square-integrable displacement fields with square-integrable weak first derivatives, i.e., $\Space{H}^1 \left( \varOmega \right) := \left\{ v : \, v \in \Space{L}^2 \left( \varOmega \right),  D v \in \Space{L}^2 \left( \varOmega \right) \right\}$. Since the weak derivative $D v$ is defined almost everywhere, no classical differentiability is required on the material interface $\varGamma^I$, which has zero volume measure.
Mathematically, $\bs{v} \in \left[ \Space{H}^1(\varOmega) \right]^3 $ allows $\grad \bs{v}_s \neq \grad \bs{v}_v$ across $\varGamma^I$, but enforces $\trace \left( \bs{v}_s \right) = \trace \left( \bs{v}_v \right)$ on $\varGamma^I$. Thus, the globally $\Space{H}^1$-conforming space enforces displacement continuity across $\varGamma^I$, while still allowing phase-wise material response through restrictions to $\varOmega_s$ and $\varOmega_v$.

The phase-wise variational form of the boundary value problem is: Find $\bs{u} \in \SpaceV{V}_0 \left( \varOmega \right)$ such that
\begin{equation}
    a \left( \bs{u},\bs{w} \right) = \ell \left( \bs{w} \right), \qquad \forall \bs{w} \in \SpaceV{V}_0 \left( \varOmega \right),
	\label{eq:weak_form}
\end{equation}
where, recalling that $\bs{u}_i = \left. \bs{u} \right|_{\varOmega_i}$ and $\bs{w}_i = \left. \bs{w} \right|_{\varOmega_i}$, bilinear and linear forms are given respectively by
\begin{align}
    a \left( \bs{u}, \bs{w} \right) & = \sum_{i \in \left\{ s, v \right\}} \int_{\varOmega_i} \bs{\varepsilon}_i \left( \bs{w}_i \right) : \bs{\sigma}_i \left(\bs{u}_i \right) \, \dd{\varOmega},
\intertext{and}
    \ell \left( \bs{w} \right) & = \sum_{i \in \left\{ s, v \right\}} \int_{\varGamma^N_i} \bs{w}_i \cdot \bar{\bs{t}}_i \, \dd{\varGamma}.
\end{align}

To solve \Cref{eq:weak_form}, we discretize $\varOmega$ using tetrahedral finite elements $e_j$, not necessarily aligned with the solid-void interface. The discretized domain is $\varOmega^h = \interior \big( \bigcup_{e_j \in \mathcal{E}} \overline{e_j} \big)$, where $\interior \left( \cdot \right)$ denotes interior and $\mathcal{E}$ is the set of all elements.
Following a Bubnov-Galerkin procedure, we choose trial and test functions from the interface-enriched generalized finite element space, defined over $\varOmega^h$, as
\begin{equation} \label{eq:enriched_space}
  \SpaceV{S}^h \left( \varOmega^h \right) = \Big\{ \bs{v}^h :  \, \bs{v}^h \left( \bs{x} \right) =  \underbrace{\sum_{\alpha \in \iota_h} N_{\alpha}(\bs{x}) \widetilde{\bs{u}}_{\alpha}}_\text{standard FEM} + \underbrace{\sum_{\beta \in \iota_w} \psi_{\beta}(\bs{x}) \widehat{\bs{u}}_{\beta}}_\text{enrichment}, \quad \widetilde{\bs{u}}_{\alpha}, \widehat{\bs{u}}_{\beta} \in \mathbb{R}^3 \Big\} \subset  \left[ \Space{H}^1 \left( \varOmega^h \right) \right]^3,
\end{equation}
so that vector-valued functions in the space that satisfy \textit{a priori} homogeneous boundary conditions are in $\SpaceV{S}_0^h := \SpaceV{S}^h \cap \SpaceV{V}_0$ (here we assume $\varOmega = \varOmega^h$). In \Cref{eq:enriched_space} $\iota_h$ is the index set of all standard mesh nodes, and thus $N_{\alpha}$ is the Lagrange shape function associated with the $\alpha$th standard node, and $\widetilde{\bs{u}}_{\alpha}$ is the corresponding displacement vector. Similarly, $\iota_w$ is the index set of all enriched nodes added along the solid-void interface, and thus $\psi_{\beta}$ and $\widehat{\bs{u}}_{\beta}$ are the enrichment function and enriched degrees of freedom (DOFs) associated with the $\beta$th enriched node.
Taking $\bs{u}^h, \bs{w}^h \in \SpaceV{S}_0^h$, \Cref{eq:weak_form} takes the form
\begin{equation}
	\sum_{e \in \mathcal{E} } \sum_{i \in \left\{ s, v \right\}} \int_{e \cap \varOmega_i} \bs{\varepsilon}_i \left( \bs{w}_i^h \right) \cdot \bs{C}_i \cdot \bs{\varepsilon}_i \left( \bs{u}_i^h \right) \, \dd{\varOmega} = \sum_{e \in \mathcal{E}} \sum_{i \in \left\{ s, v \right\}}  \int_{\partial e \cap \varGamma_i^N} \bs{w}_i^h \cdot \bar{\bs{t}}_i \, \dd{\varGamma},  \quad \forall \bs{w}^h \in \SpaceV{S}_0^h,
\end{equation}
where Voigt notation (engineering strain) is implied following standard FEM procedures, and thus $\bs{C}_i$ denotes the constitutive matrix. 

While the numerical quadrature of local arrays in elements not cut by $\varGamma^I$ follows standard FEM procedures, cut elements are subdivided into \textit{integration elements} that are aligned with the interface. Across the interface, $\bs{v}^h$ is $\Space{C}^0$-continuous but has discontinuous gradients; within each integration element, the approximation is smooth. 
Therefore, exact integration of the stiffness contribution is attained with one quadrature point per affine integration tetrahedron, under constant material properties.

For the $k$th integration element, the subscript $i \in \left\{ s, v \right\}$ is selected according to the phase occupied by that integration element.
Numerical quadrature in each integration element is carried out using an iso-parametric mapping. Therefore, the master element $ \hat{e} = \left\{ \bs{\xi} = \left( \xi, \eta, \zeta \right) \in \mathbb{R}^3 : \, \xi \geq 0, \eta \geq 0, \zeta \geq 0, \xi + \eta + \zeta \leq 1 \right\}$, is mapped to the physical integration element associated with $k$ using the same linear shape functions of the linear tetrahedron. For the surface integral we also use a mapping from the master triangular element $\hat{t} = \left\{ \bar{\bs{\xi}} = \left( \bar{\xi}, \bar{\eta} \right) \in \mathbb{R}^2 : \, \bar{\xi} \geq 0, \bar{\eta} \geq 0, \bar{\xi} + \bar{\eta} \leq 1 \right\}$.

Denoting by $\iota_{e}$ the index set of integration elements associated with the parent element $e$(i.e., $k \in \iota_{e}$), and adopting an iso-parametric formulation, the stiffness matrix and force vector are computed, respectively, as
\begin{align}
    \bs{k}_{e} & = \sum_{k \in \iota_{e}} \int_{ \hat{e} } \bs{B}_k^\intercal \bs{C}_{i (k)} \bs{B}_k \, j_{\hat{e}} \, \dd{ \bs{\xi} }, \label{eq:ke}
\intertext{and}
	\bs{f}_e & = \sum_{k \in \iota_{e}} \int_{ \hat{\varGamma}_i^N} \bs{\Phi}_k^\intercal \bar{\bs{t}}_{i (k)} \, j_{\hat{t}} \, \dd{ \bar{\bs{\xi}} },
    \label{eq:fe}
\end{align}
where we made explicit that the subscript $i$ is chosen depending on the phase where the $k$th integration element lies, so $i (k)=s$ if the integration element lies in $\varOmega_s$, and $i (k)=v$ if it lies in $\varOmega_v$. In \Cref{eq:ke} $j_{\hat{e}} =  \left| \det \bs{J}_k \right|$ with $\bs{J} = \grad_{\bs{\xi}} \bs{x}^\intercal$ denoting the Jacobian matrix of the tetrahedron geometry mapping. For the surface integral in \Cref{eq:fe}, $\hat{\varGamma}_i^N$ is the preimage of the Neumann portion of the boundary of the kth integration element, and $j_{\hat{t}}$ the positive surface Jacobian.

In ~\Cref{eq:fe} $\bs{\Phi}_k = \bs{\varphi} \otimes \bs{I} $, where $\bs{\varphi} =  \begin{bmatrix} N_1 & N_2 & \cdots & \psi_1 & \cdots \end{bmatrix}$ collects the scalar shape and enrichment functions in a row vector, $\otimes$ is the Kronecker product and $\bs{I}$ the $3 \times 3$ identity matrix. The strain-displacement matrix $\bs{B}_k$ is computed as  $\bs{B}_k \equiv \DifferentialOperator \bs{\varphi} := \begin{bmatrix}
		\DifferentialOperator N_1 & \DifferentialOperator N_2 & \cdots & \DifferentialOperator \psi_1 & \cdots \end{bmatrix} $, with the differential operator $\DifferentialOperator$ defined in Voigt notation as
\begin{equation} \label{eq:differential_operator}
  \DifferentialOperator \equiv
  \begin{bmatrix}
    \pdv{}{x} & 0 & 0 & 0 & \pdv{}{z} & \pdv{}{y} \\
    0 & \pdv{}{y} & 0 & \pdv{}{z} & 0 & \pdv{}{x} \\
    0 & 0 & \pdv{}{z} & \pdv{}{y} & \pdv{}{x} & 0 \\
  \end{bmatrix}^\intercal.
\end{equation}
It is worth noting that there are two mappings involved in Equations~\Cref{eq:ke} and \Cref{eq:fe} since the Lagrange shape functions $N_{\alpha}$ use a different iso-parametric mapping than that used for integration.
An integration point is first mapped from the integration master element to the physical integration element. The resulting physical coordinate is then mapped back to the parent element $e$ to evaluate the standard shape functions $N_{\alpha}$. The enrichment functions $\psi_{\beta}$ are evaluated using the integration-element mapping. Consequently, the derivatives in $\bs{B}_k$ involve different Jacobians for $N_\alpha$ and $\psi_\beta$; in both cases, the chain rule is applied to obtain derivatives with respect to the physical coordinates. Finally, the constitutive 3-D matrix in Voigt notation is
\begin{equation}
  \bs{C}  = \frac{E}{\left(1+\nu\right)\left(1-2\nu\right)} \begin{bmatrix}
  	\left(1 - 2 \nu\right) \bs{I} + \nu \bs{1} & \bs{0} \\
  	\bs{0} & \left(\frac{1-2\nu}{2}\right) \bs{I} \\
  \end{bmatrix}
\end{equation}
where $\bs{I}$ is the $3 \times 3$ identity matrix and $\bs{1}$ is the $3 \times 3$ matrix of ones. Implementation details can be found in Aragón and Duarte~\cite{aragon2023fundamentals}.

Once all local arrays are computed, the global stiffness matrix $\bs{K}$ and the global force vector $\bs{F}$ are obtained as
\begin{equation}
	\bs{K} = \assembly_{e \in \mathcal{E}} \bs{k}_e, \qquad \bs{F} = \assembly_{e \in \mathcal{E}} \bs{f}_e,
\end{equation}
where $\assembly{}$ is the standard finite element assembly operator.
The global degree of freedom vector is then obtained by solving the system of linear equations $\bs{K} \bs{U} = \bs{F}$.
For more detailed explanations on IGFEM, the reader is referred to \cite{Soghrati:2012, aragon2020stability, aragon2023fundamentals}, and references therein.

\subsection{Boundary conditions in immersed analysis}
\label{sec:boundcond}

\begin{figure}[t]
  \centering
  \begin{subfigure}[t]{0.48\linewidth}
    \centering
    \includegraphics[]{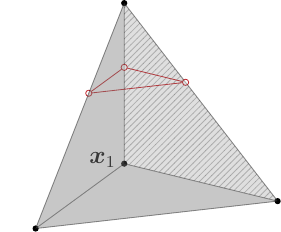}
    \caption{}
    \label{fig:dirichlet}
  \end{subfigure}
  \begin{subfigure}[t]{0.48\linewidth}
    \centering
  	\includegraphics[]{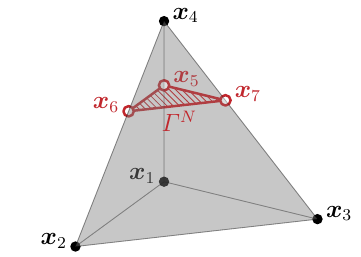}
    \caption{}
    \label{fig:neumann}
  \end{subfigure}
  
  \caption{(a) A Dirichlet boundary condition is prescribed on a face of a tetrahedral element (shaded area defined by nodes $\bs{x}_1$, $\bs{x}_3$, and $\bs{x}_4$). The tetrahedron is cut by a solid-void interface $\varGamma^I$, thereby adding enriched nodes $\bs{x}_5$, $\bs{x}_6$, and $\bs{x}_7$. Because $\bs{x}_5$ and $\bs{x}_7$ also belong to the Dirichlet region $\varGamma^D$, the field is also prescribed at these enriched nodes. Enriched DOFs are determined by solving~\Cref{eq:localDOF} and they are prescribed strongly thereafter; and (b) A traction is prescribed on the solid-void interface $\varGamma^N$ defined by nodes $\bs{x}_5$, $\bs{x}_6$, and $\bs{x}_7$ (red shaded area). The computation of the force vector given by \Cref{eq:fe} requires a modified definition of the Jacobian matrix.}
\end{figure}

One advantage of IGFEM over other enriched finite element formulations is that, similarly to standard FEM, the original mesh nodes retain their physical meaning and associated degrees of freedom (DOFs)~\cite{Soghrati:2012}. Consequently, Dirichlet boundary conditions can be prescribed strongly on enriched discretizations after determining the values of enriched DOFs through the solution of local problems~\citep{van2019stable,zhang2021on}. For example, consider the cut tetrahedral element shown in \Cref{fig:dirichlet}. A Dirichlet boundary condition is prescribed on the shaded face defined by nodes $\bs{x}_1$, $\bs{x}_3$, and $\bs{x}_4$, while the element is intersected by a material interface (red), introducing enriched nodes $\bs{x}_5$, $\bs{x}_6$, and $\bs{x}_7$. Since nodes $\bs{x}_5$ and $\bs{x}_7$ also lie on the Dirichlet boundary, the corresponding enriched DOFs must satisfy the prescribed displacement field. Their values     follow from the local interpolation problem (cf.~\Cref{eq:enriched_space}):
\begin{equation} \label{eq:localDOF}
\begin{split}
  \widehat{\bs{u}}_5 &= \bar{\bs{u}}(\bs{x}_5) - N_1(\bs{x}_5) \widetilde{\bs{u}}_1 - N_4(\bs{x}_5) \widetilde{\bs{u}}_4,\\
  \widehat{\bs{u}}_7 &= \bar{\bs{u}}(\bs{x}_7) - N_3(\bs{x}_7)\widetilde{\bs{u}}_3 - N_4(\bs{x}_7)\widetilde{\bs{u}}_4,
\end{split}
\end{equation}
where $\widetilde{\bs{u}}_i=\bar{\bs{u}}(\bs{x}_i), \, i \in \left\{ 1, 3, 4 \right\}$. Once known, $\widehat{\bs{u}}_5$ and $\widehat{\bs{u}}_7$ are prescribed as in standard FEM.

Neumann boundary conditions are prescribed through the standard finite-element boundary integral in \Cref{eq:fe}. As an illustrative example, consider the traction applied on the immersed interface shown in \Cref{fig:neumann}. The interface is represented by a triangular surface element connecting enriched nodes $\bs{x}_5$, $\bs{x}_6$, and $\bs{x}_7$, over which the boundary integral is evaluated. The corresponding Jacobian matrix is given by
\begin{equation} \label{eq:StandardJac}
	\bs{J} = \bs{X}_{\vartriangle}^\intercal \pdv{\bs{N}_{\vartriangle}}{\bs{\xi}_{\vartriangle}} = 
	\begin{bmatrix}
		\pdv{x}{\xi} & \pdv{x}{\eta} \\
  \pdv{y}{\xi} & \pdv{y}{\eta} \\
  \pdv{z}{\xi} & \pdv{z}{\eta}
	\end{bmatrix},
\end{equation}
where $\bs{X}_{\vartriangle}$ contains the coordinates of nodes $\bs{x}_5$, $\bs{x}_6$, and $\bs{x}_7$, and $\partial\bs{N}_{\vartriangle}/\partial\bs{\xi}_{\vartriangle}$ denotes the derivatives of the triangular shape functions with respect to the local coordinates $\left( \xi,\eta \right)$ of the master triangular element. Since the surface element is embedded in three-dimensional space, the Jacobian in \Cref{eq:StandardJac} is a $3\times2$ matrix. Consequently, its determinant and inverse are not defined, preventing the direct application of the standard finite-element integration procedure. To overcome this difficulty, we follow the approach of~\citet{devloo1997pz}, who presents a generalization of the Jacobian when dealing with lower-dimensional elements. Furthre details are given in \Cref{Ap:devloo}.

It is worth mentioning that it is possible to have fully-immersed Dirichlet boundaries (for instance, consider the material interface $\varGamma^N$ in \Cref{fig:neumann} changes to $\varGamma^D$). In such cases, a multi-point constraint system is setup and Dirichlet boundary conditions can still be prescribed strongly~\citep{van2019stable}.

\section{Examples of citations, figures, tables, references}
\label{sec:form}

\subsection{Topology description}
In this work, the topology is represented by the zero contour of a level set function, following the approach originally proposed by \citet{sethian2000structural}. To wit,
\begin{equation}
	\begin{cases}
		\phi(\bs{x})=0, & \text{if } \bs{x} \in \varGamma^I,\\
		\phi(\bs{x})<0, & \text{if } \bs{x} \in \varOmega_s ,\\
		\phi(\bs{x})>0, & \text{if } \bs{x} \in \varOmega_v.
	\end{cases}
\end{equation}

Compactly supported radial basis functions (RBFs) are employed to interpolate the level set function, and the centers of these RBFs are placed in a structured grid (which does not necessarily coincide with the locations of the nodes of the finite element discretization).
The level set function $\phi(\bs{x})$ is therefore expressed as
\begin{equation}
	\phi \left( \bs{x} \right) = \sum_{i \in \iota_r} \vartheta_i \left( \bs{x}, \bs{x}_i \right) s_i,
	\label{eq:level_set}
\end{equation}
where $\iota_r$ is the index set of all RBFs, $\bs{x}_i$ the coordinate of the center of the $i$th RBF, and $s_i$ its associated coefficient. For our compactly supported RBFS we use fourth-order Wendland functions~\citep{wendland1995piecewise}, defined as
\begin{align}
  \vartheta_i \left( \bs{x}, \bs{x}_i \right) 
  = 
  \begin{cases}
  \left( 1 - r_i \left( \bs{x}, \bs{x}_i \right) \right)^4 \left( 4 r_i \left( \bs{x}, \bs{x}_i \right) + 1 \right) & \text{with } 0\leq r_i \left( \bs{x}, \bs{x}_i \right) = \frac{\norm{\bs{x} - \bs{x}_i}}{r_s} \leq 1,\\
  0 & \text{else}
  \end{cases}
  \label{eq:wend}
\end{align}

By defining the matrix $\bs{\vartheta} \left( \bs{x} \right)^\intercal = \begin{bmatrix} \vartheta_1(\bs{x}, \bs{x}_1) &  \vartheta_2(\bs{x}, \bs{x}_2) & \cdots & \vartheta_{r}(\bs{x}, \bs{x}_r) \end{bmatrix} \in \mathbb{R}^r$, where $r = \left| \iota_r \right|$ ($\left| \, \cdot \, \right|$ denotes set cardinality), and the coefficient vector $\bs{s} = \begin{bmatrix} s_1 &  s_2  & \cdots & s_{r} \end{bmatrix}^\intercal \in \mathbb{R}^r$, \Cref{eq:level_set} can be written as 
\begin{equation}
\phi \left( \bs{x} \right)  = \bs{\vartheta} \left( \bs{x} \right)^\intercal \bs{s}.
\label{eq:DesToLSF}
\end{equation}
Note that in our framework, $\bs{s}$ also represents our design variable vector. As it is traditionally done, the level set function is computed at every node of the finite element mesh; denoting by $\bs{\phi}$ such vector, we then compute it as
\begin{equation}
  \bs{\phi} = \bs{\Theta} \bs{s},
\end{equation}
where $\bs{\Theta} \in \mathbb{R}^{r \times r} $ is a sparse matrix (given the local support of the RBFs) that is computed only once at the beginning of the simulation. 
The matrix $\bs{\Theta}$ is assembled once, with each row containing the RBF evaluations that determine the influence of the coefficient vector $\bs{s}$ on the corresponding nodal value of the level-set field.

New enriched nodes are introduced at the intersections between the zero level set and the edges of finite elements in the original mesh. Since the exact location of these intersections is generally unknown, it is approximated by interpolating the level set values at the two end nodes of intersected element edges. This approximation relies on the fact that a continuous function that changes sign over an interval must possess at least one root within that interval. Accordingly, consider two mesh nodes $\bs{x}_i$ and $\bs{x}_j$ satisfying  $\phi \left( \smash{\bs{x}_i} \right) \phi \left( \smash{\bs{x}_j} \right) < 0$. In previous topology optimization works that used IGFEM~\cite{van2021interface}, the location of the enriched node $\bs{x}_n$ was approximated via the linear interpolation
\begin{equation}
	\bs{x}_n = \bs{x}_i - \frac{\phi_i}{\phi_j - \phi_i} \left( \bs{x}_j - \bs{x}_i \right),
\end{equation}
where $\phi_j \equiv \phi(\bs{x}_j)$ and $\phi_i \equiv \phi(\bs{x}_i)$.

In the present study, we also consider a quadratic interpolation of the level set to obtain a higher-order approximation of the intersection location. Consider the bijective mapping
\begin{equation} \label{eq:bijective_mapping}
  \bs{x} \left( \xi \right) = \frac{1-\xi}{2}\bs{x}_i + \frac{1+\xi}{2}\bs{x}_j \qquad \Rightarrow \qquad \xi \left(\bs{x}\right) = 2 \frac{ (\bs{x} - \bs{x}_i ) \cdot (\bs{x} - \bs{x}_j ) }{\norm{\smash{\bs{x}_j} - \bs{x}_i}^2 } -1.  
\end{equation}
where $\xi$ is the one-dimensional master coordinate satisfying $\xi \in [-1,1]$ along the edge from $\bs{x}_i$ to $\bs{x}_j$.
Using this normalized coordinate, the level set function is approximated along the edge by the quadratic interpolation
\begin{equation}
\begin{split}  
  \arc{\phi} \left(\xi\right) & = \frac{\xi (\xi-1)}{2}\phi_i + (1 - \xi^2) \phi_k + \frac{\xi(\xi+1)}{2}\phi_j, \\
   & = \underbrace{\frac{1}{2} \left( \phi_i+\phi_j -2 \phi_k \right)}_{a} \xi ^2 + \underbrace{\frac{\left(\phi_j - \phi_i \right)}{2}}_{b} \xi + \underbrace{\phi_k,}_{c}
\end{split}
\end{equation}
where $\phi_k = \phi(\bs{x}_k)$ denotes the level set value at the midpoint $\bs{x}_k = \tfrac{1}{2}\left( \smash{\bs{x}_j + \bs{x}_i} \right)$ along the edge.
This quadratic interpolation is exactly matches the level set function at the three interpolation points, i.e., $\arc{\phi} \left( -1 \right) = \phi_i$, $\arc{\phi} \left( 0 \right) = \phi_k$, and $\arc{\phi} \left( 1 \right) = \phi_j$.
Accordingly, the location of the enriched point $\bs{x}_n$ is obtained by determining the root of the quadratic interpolation, i.e., by solving $\phi_q(\xi)=0$. Solving the resulting quadratic equation yields
\begin{equation}
    \xi_n = \frac{-b \pm \sqrt{b^2 - 4ac}}{2a} \in \left( -1, 1 \right).
\label{eq:xiE}
\end{equation}
The enriched-point coordinates are then obtained via \Cref{eq:bijective_mapping} as $\bs{x} \left( \smash{\xi_n} \right)$. 

\begin{figure}[t]
  \centering
    \includegraphics{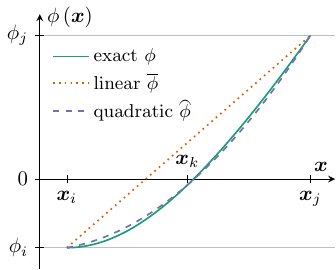}
  \caption{Comparison of different interpolations of the level set function. The curve labeled ``exact $\phi$'' represents the actual nonlinear level-set function along the edge defined by nodes $\bs{x}_i$ and $\bs{x}_j$, while the dotted and dashed curves denote the linear and quadratic interpolations, respectively.}
    \label{fig:interpolation}
\end{figure}

The difference between linear and quadratic interpolations is schematically illustrated in \Cref{fig:interpolation}. By incorporating the value at the midpoint, the quadratic interpolation better approximates the nonlinear level-set field, thereby improving the representation of the zero contour and, consequently, of the material interface.

\subsection{Problem formulation}
\label{ssec:formulation}

The optimization problem is given by
\begin{equation}\label{eq:TOform}
\begin{split}
  \bs{s}^\star = \argmin_{\bs{s} \in \SpaceV{D}} \quad & \Phi \left( \bs{s} \right) \\
    \text{subject to } \quad & \bs{K} \bs{U}=\bs{F}, \\
    & V_s \leq V_c, \\
    &-1 \leq s_i \leq 1, \quad i = \left\{ 1, \ldots, r \right\},
\end{split}
\end{equation}
where $\Phi \left( \bs{s} \right)$ is the objective function, $V_s$ is the volume of solid, and $V_c$ is the maximum allowed solid volume. The box constraint on the design variables prevents the level set function from becoming too steep~\cite{Boom:2021aa}.

To minimize the energy release rates throughout the boundary, at the locations of all potential cracks (which correspond to the location of all enriched nodes), we use the $p$-mean aggregation function
\begin{equation}
    \Phi = \left( \frac{1}{ \abs{\iota_w} } \sum_{i \in \iota_w} G_i^p \right)^{\frac{1}{p}},
    \label{eq:chJ}
\end{equation}
where $G_i$ is the ERR associated to the $i$th potential crack (at the $i$th enriched node location), and $p$ is the aggregation parameter.
The $p$-mean aggregation therefore combines the ERRs associated with all potential cracks into a single differentiable objective function.

\begin{figure}[t]
 \centering
 \includegraphics{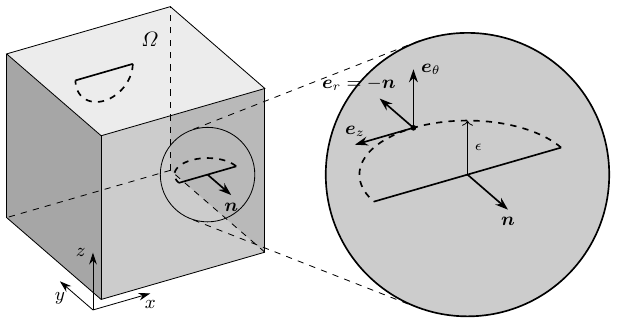}
\caption{Assumed crack nucleation model. Half penny-shaped cracks are assumed to nucleate perpendicularly to the boundary. The inset shows the local coordinate system adopted for evaluating the stress intensity factors and consequently the energy release rate, where $\bs{e}_r$ is a unit vector opposite to the boundary normal, $\bs{e}_{\theta}$ the crack opening direction, and $\bs{e}_z$ is orthogonal to both.}
  \label{plot:crackeddomain}
\end{figure}

\begin{figure}[t]
\centering
\includegraphics{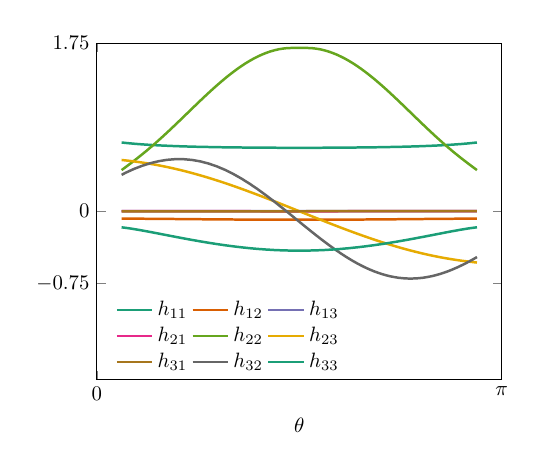}
\caption{Weight functions $h_{ij}$ used for the evaluation of stress intensity factors via~\Cref{eq:3DWeightFunction}. }
\label{plot:H}
\end{figure}

Considering half penny-shaped cracks, the ERR is related to stress intensity factors via
\begin{equation} \label{eq:3DERR}
    G = \pi \epsilon \left( \frac{\overline{\SIFmodeI}^2+\overline{\SIFmodeII}^2}{\bar{E}} + \frac{\overline{\SIFmodeIII}^2}{2 \mu_L} \right),
\end{equation}
where $\epsilon$ is the crack radius, $\mu_L = E/\left(2(1+\nu)\right)$ is the shear modulus, $\bar{E} = E/(1-\nu^2)$ is the effective Young's modulus under plane strain, with $E$ and $\nu$ denoting Young's modulus and Poisson's ratio, respectively. The quantities $\overline{\SIFmodeI}$, $\overline{\SIFmodeII}$, and $\overline{\SIFmodeIII}$ denote the stress  intensity factors for fracture modes I, II, and III, respectively, normalized according to $\overline{\bs{K}}=\dfrac{\bs{K}}{\sqrt{\pi\epsilon}}$, and are evaluated as~\citep{alidoost2020energy}
\begin{equation}
    \begin{bmatrix}
        \overline{\SIFmodeI} \\
        \overline{\SIFmodeII} \\
        \overline{\SIFmodeIII}
    \end{bmatrix}
    =
    \underbrace{\begin{bmatrix}
        h_{11} & h_{12} & h_{13} \\
        h_{21} & h_{22} & h_{23} \\
        h_{31} & h_{32} & h_{33}
    \end{bmatrix}}_\text{$\bs{H}$}
    \underbrace{\begin{bmatrix}
        \sigma_{\theta \theta} \\
        \sigma_{r \theta} \\
        \sigma_{z \theta}
    \end{bmatrix}}_{\bs{\sigma}'},
    \label{eq:3DWeightFunction}
\end{equation}
where the stress field is given in cylindrical coordinates $\left(r, \theta, z \right) $ (see \Cref{plot:crackeddomain}), and  $\bs{H}$ denotes the matrix of weight functions. These functions, which are shown in \Cref{plot:H}, are computed herein numerically using the finite element domain integral method of \citet{nagai2013stress} (see \Cref{Ap:weight}).

Substituting \Cref{eq:3DWeightFunction} into \Cref{eq:3DERR} yields
\begin{equation}
    \label{eq:befERRfinal}
    G = \frac{\pi \epsilon}{2 \mu_L \bar{E}} \left( \bs{H}' \bs{\sigma}' \right) ^\intercal   \bs{H}' \bs{\sigma}'.
\end{equation}
where
\begin{equation}
    \bs{H}' = \begin{bmatrix}
        \sqrt{2\mu_L}h_{11} & \sqrt{2\mu_L}h_{12} & \sqrt{2\mu_L}h_{13} \\
        \sqrt{2\mu_L}h_{21} & \sqrt{2\mu_L}h_{22} & \sqrt{2\mu_L}h_{23} \\
        \sqrt{\bar{E}}h_{31} & \sqrt{\bar{E}}h_{32} & \sqrt{\bar{E}}h_{33}
    \end{bmatrix}.
\end{equation}

The stress field used in the ERR evaluation depends on the assumed crack orientation. Following the work of \citet{zhang2021on} for two-dimensional structures, herein cracks are assumed to nucleate perpendicularly to the solid boundary, as illustrated in \Cref{plot:crackeddomain}. In the three-dimensional setting, however, this condition alone does not uniquely define the crack orientation, since infinitely many planes satisfy the perpendicularity condition. We resolve this remaining degree of freedom using the stress state of the undamaged (uncracked) domain at the nucleation point: among all admissible planes, the crack is oriented so that its normal aligns with the direction of maximum principal (tensile) stress, i.e., so that the resolved normal stress on the crack plane is maximized. Once the crack is oriented in this way, the associated local coordinate system $(r,\theta,z)$ is constructed (\Cref{plot:crackeddomain}). This orientation is equivalently seen to maximize the hoop stress $\sigma_{\theta\theta}$ ahead of the crack front ($\theta = 0$) in that frame; this is consistent with the maximum-hoop-stress (maximum tangential stress) hypothesis of \citet{erdogan1963crack}, originally used to predict the kink direction of an existing crack, but which we adapt here to select the nucleation orientation of a virtual crack embedded in an otherwise smooth stress field. Because a negative mode-I stress intensity factor corresponds to interpenetration of the crack faces rather than physical opening, only crack configurations with $\SIFmodeI > 0$ are retained in the ERR evaluation and the subsequent aggregation in \Cref{eq:chJ}; the corresponding transformed stress components in the local crack coordinate system are then used to evaluate the ERR.

With these assumptions, the ERR can finally be expressed directly in terms of the stress tensor $\bs{\sigma}$ in the global Cartesian coordinate system as
\begin{equation}
    \label{eq:ERRfinal}
    G = \frac{\pi \epsilon}{2 \mu_L \bar{E}} \bs{Q}_R^\intercal \bs{\sigma}^\intercal \bs{F}\bs{\sigma} \bs{Q}_R,
\end{equation}
where $\bs{Q}_R$ and $\bs{F}$ account for the transformation between the global Cartesian coordinate system and the local crack coordinate system, as well as the contribution of the weight functions. Additional details regarding the derivation and construction of these matrices are provided in~\Cref{Ap:ERR_comp}.

\subsection{Evaluation of Stress}
\label[section]{sec:sip}

It is well known that enriched formulations may overestimate the stress field locally in cut elements. Therefore, a stress improvement procedure (SIP) is used herein to obtain a smoother more accurate stress field than that provided by directly calculated stresses from FEA~\citep{payen2012stress, sharma2018improved, zhang2021on}. The method is based on the Hu-Washizu variational principle~\citep{chan1968variational} and was originally proposed by \citet{payen2012stress}. It was subsequently extended to three-dimensional problems by \citet{sharma2018improved} and, more recently, to enriched and unfitted finite element formulations by \citet{zhang2022improved}.

\begin{figure}[t]
	\centering
	\begin{subfigure}[b]{0.4\textwidth}
	    \centering
		\includegraphics[width=0.6\linewidth]{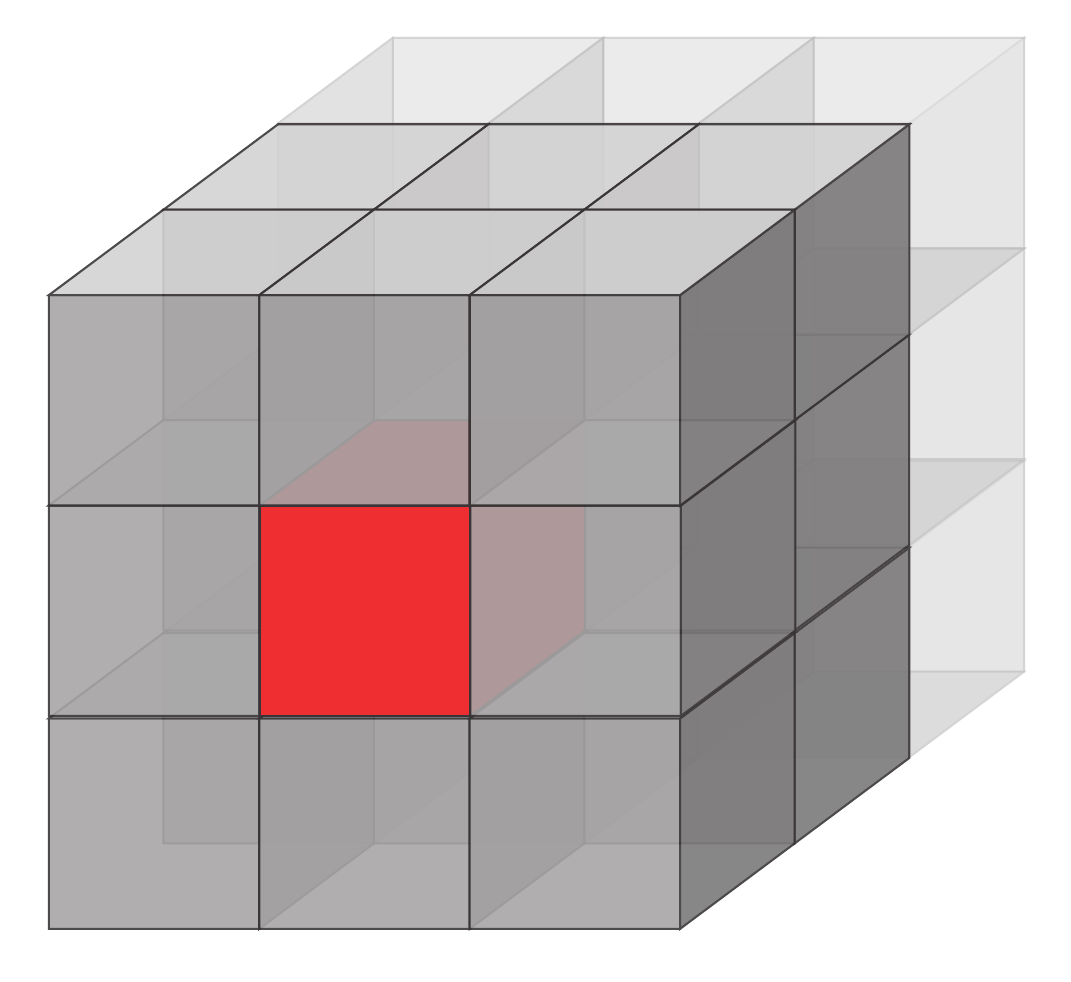}
		\caption{}
		\label{fig:element_stress}
	\end{subfigure}
    \begin{subfigure}[b]{0.4\textwidth}
        \centering
	    \includegraphics[width=0.6\linewidth]{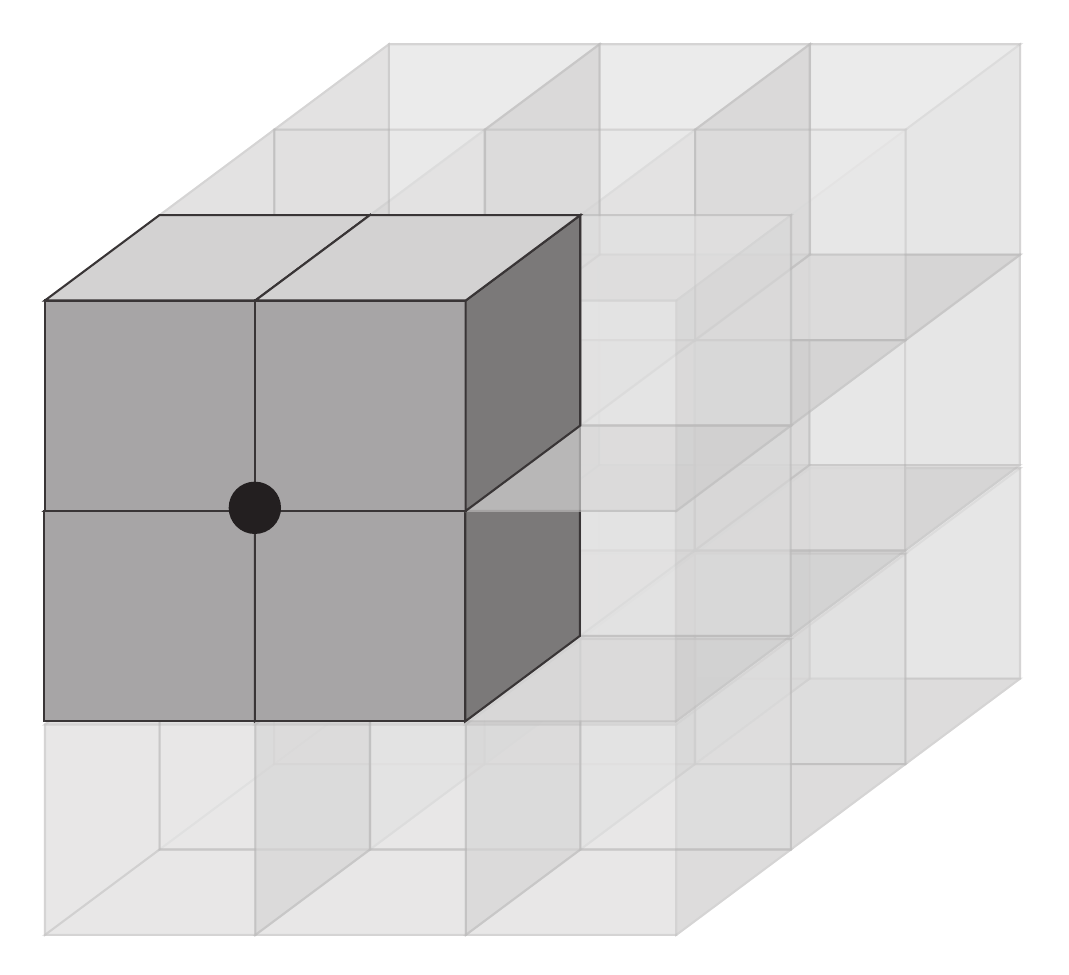}
	    \caption{}
	    \label{fig:nodal_stress}
    \end{subfigure}
    \caption{Schematic of the calculation domain for the improved stress computation: (a) Target element (red) and neighboring elements (dark gray) jointly define the element patch to calculate the improved stress field within the target element; and (b): Target surface node (black) and its element patch, used to recover and averaged nodal stress value.}
\end{figure}

The recovered stress field in a target element is obtained by fitting a quadratic polynomial over a patch $\mathcal{P}$ consisting of the element and its neighboring elements. An example of such a computational patch is shown in \Cref{fig:element_stress}, where the stress is recovered for the element highlighted in red using its stress field and those of the neighboring elements, shown in dark gray.  The polynomial coefficients of the fit are determined by simultaneously enforcing equilibrium in a weak sense over the patch and minimizing the discrepancy with respect to directly-calculated finite element stresses. These conditions lead to the following system of equations:
\begin{equation}
\label{eq:stress}
	\begin{bmatrix}
	\sum_{ e \in \mathcal{P}} \begin{pmatrix}
			\int_e \bar{\bs{E}}_{\sigma}^\intercal \bs{E}_{\sigma} \, \dd{e} \\
			\int_e \bs{E}_{\zeta}^\intercal \bs{\partial}_{\sigma} \bs{E}_{\sigma} \, \dd{e}
	\end{pmatrix}
    \end{bmatrix}
    \hat{\bs{\sigma}}
    =
    \begin{Bmatrix}
    	\sum_{e \in \mathcal{P}} \begin{pmatrix}
    		\int_e \bar{\bs{E}}_{\sigma}^\intercal \bs{\sigma}_e^h \, \dd{e} \\
    		-\int_e \bs{E}_{\zeta}^\intercal \bs{b} \, \dd{e}
    	\end{pmatrix}
    \end{Bmatrix},
\end{equation}
where $\hat{\bs{\sigma}}$ denotes a $60 \times 1$ vector of stress polynomial coefficients, $\bs{\sigma}_e^h$ the directly-calculated stress from the finite element solution, $\bs{b}$ is the body force, $\bar{\bs{E}}_{\sigma}$, $\bs{E}_{\sigma}$, $\bs{E}_{\zeta}$ are interpolation matrices, and $\bs{\partial}_{\sigma}$ is a differential operator. The definitions of these matrices and operators are provided in \Cref{Ap:matrices}. Note that in this work we prescind of body forces, so $\bs{E}_{\zeta}^\intercal \bs{b} = \bs{0}$.
After solving \Cref{eq:stress}, the recovered improved stress within the target element $e$ is
\begin{equation}
 \bs{\sigma}_e = \bs{E}_{\sigma} \hat{\bs{\sigma}}.
 \end{equation}
Repeating this procedure for every element yields a recovered stress field defined element-wise throughout the mesh. A continuous nodal stress field is then obtained by averaging, at each node, the recovered stresses of all elements sharing that node---denoted as $\mathcal{P}_i$. The stress at the $i$th node is
\begin{equation}
	\bs{\sigma} \left( \bs{x}_i \right) = \frac{1}{\left| \mathcal{P}_i \right|} \sum_{e \in \mathcal{P}_i} \bs{\sigma}_e \left( \bs{x}_i \right)
	\label{eq:recovered_stress},
\end{equation}
where $\bs{\sigma}_e \left( \bs{x}_i \right) = \bs{E}_{\sigma} \left( \bs{x}_i \right) \bs{\hat{\sigma}}_e$ is the enhanced element stress of $e \in \mathcal{P}_i$ evaluated at node $\bs{x}_i$. A nodal patch is also illustrated in \Cref{fig:nodal_stress}. For more details on the stress improvement procedure, the reader is referred to \Cref{Ap:matrices} and works elsewhere~\citep{payen2012stress, sharma2018improved, zhang2021on,zhang2022improved}.

\subsection{Sensitivity analysis}
\label{sec:sensitivity_analysis}

An adjoint Lagrangian function $ \Lambda = \Phi + \bs{\lambda}^\intercal \left( \bs{KU}-\bs{F} \right)$ is established with the adjoint vector $\bs{\lambda}$. The sensitivity of the objective function is then expressed as the derivative of $\Lambda$ with respect to the $j$th design variable $s_j$:
\begin{equation}
	\begin{split}
		\pdv{\Lambda}{s_j} &=  \pdv{\Phi}{s_j} + \pdv{\Phi}{\bs{U}} \pdv{\bs{U}}{s_j} + \bs{\lambda}^\intercal \left( \pdv{\left( \bs{KU} \right)}{s_j}  - \smash{ \cancel{\pdv{\bs{F}}{s_j}} } \right) \\
		&=  \pdv{\Phi}{s_j} + \left( \pdv{\Phi}{\bs{U}} + \bs{\lambda}^\intercal \bs{K} \right) \pdv{\bs{U}}{s_j} + \bs{\lambda}^\intercal  \pdv{\bs{K}}{s_j} \bs{U}. \\
	\end{split}
	\label{eq:lag}
\end{equation}
Since the external loads are prescribed exclusively on the non-design domain, the load vector does not depend on the design variables. Hence, $\pdv{\bs{F}}{s_j} = 0$, and the adjoint equation 
\begin{equation}
	\frac{\partial \Phi}{\partial \bs{U}} + \bs{\lambda}^\intercal \bs{K} = \bs{0},
	\label{eq:chadj}
\end{equation}
is solved to determine the adjoint vector $\bs{\lambda}$.
The concrete forms of  $\partial \Phi/ \partial s_j$, $\partial \Phi / \partial \bs{U}$, and $\partial \bs{K}/ \partial s_j$ are given in \Cref{Ap:sens}.
Furthermore, analytical sensitivities have been verified by means of finite differences, results that are summarized in \Cref{Ap:sensVer}.

\subsection{Regularization of the LSF}
A level set function is often initialized as a signed-distance function, for which $\left| \grad \phi \right| = 1$. During optimization, however, the LSF generally loses this distance property: as the design evolves, the field may become locally too steep or too flat, distorting the scalar representation \cite{osher2004level,allaire2004structural,van2013level}. Such distortion can slow convergence and degrade the geometric quality of the zero level set, i.e., the material interface.

The conventional remedy is to re-initialize $\phi$ by solving a Hamilton--Jacobi PDE at pseudo-steady state~\cite{andrew2000level, osher2004level}, which requires an additional PDE solve at every re-initialization step and adds non-negligible cost to the optimization loop. In this work, we instead adopt a solution inspired by the work of Yamasaki et al.~\cite{yamasaki2010structural}, which re-establishes $\phi$ as a signed-distance function directly, without solving an additional PDE. The nodal level set values are recomputed as the signed geometric distance from each node to the reconstructed material interface. This is done by orthogonally projecting each node onto the piecewise-linear representation of the zero level set, or, when no such projection exists, by selecting the nearest enriched node instead. Because the distance property is enforced directly at the nodal level, $\left|\grad \phi\right| = 1$ is recovered approximately in the discretized domain.

\subsection{Topological derivatives}
\label{ch:TD}

In level-set-based topology optimization, the evolution of the design is governed by the motion of the existing material interfaces. As a consequence, the optimized topology may exhibit a considerable dependence on the initial configuration, in particular on the number and spatial distribution of the initially prescribed holes. To alleviate this dependence, the shape-sensitivity-driven evolution is complemented here by a topological derivative criterion for the nucleation of new holes.

The topological sensitivity is evaluated using the three-dimensional Neumann asymptotic expression derived by \citet{garreau2001topological}, corresponding to the nucleation of an infinitesimal traction-free cavity. The resulting topological derivative field is given by
\begin{equation}
  \mathcal{T} \left( \bs{x} \right) =
  -\frac{\pi\left(\lambda_L + 2\mu_L \right)}{\mu_L \left( 9\lambda_L + 14\mu_L   \right)}
  \left[
  20 \mu_L \, \bs{\sigma} \left( \bs{u} \right) : \bs{\epsilon} \left( \bs{\lambda} \right)
  +
  \left( 3\lambda_L - 2\mu_L \right) \trace \bs{\sigma} \left( \bs{u} \right)
  \trace \bs{\epsilon}\left( \bs{\lambda} \right)
  \right],
  \label{eq:TD_3D}
\end{equation}
where $\lambda_L$ and $\mu_L$ are the Lamé parameters, and $\bs{\lambda}$ is the adjoint field associated with the ERR-based objective function (see \Cref{sec:sensitivity_analysis}).

Although the topological derivative provides information regarding favorable locations for hole nucleation, practical implementations generally require additional admissibility criteria to control the introduction of new holes. \citet{allaire2005structural} noted that the insertion of finite-sized holes based solely on the sign of the topological derivative may lead to undesirable topological changes, and therefore introduced additional criteria to regulate the nucleation process. In the present work, the topological derivative is evaluated throughout the domain according to \Cref{eq:TD_3D}, and a new hole is allowed to nucleate only if the candidate location satisfies the following admissibility criterion:
\begin{equation}
\phi(\bs{x}) < -4h ,
\label{eq}
\end{equation}
where $h$ denotes the mesh size.
Finally, we evaluate the topological derivative criterion to nucleate new holes every five optimization iterations. This allows sufficient boundary evolution between successive nucleation attempts while avoiding premature convergence to a local optimum that may hinder the introduction of beneficial cavities~\cite{allaire2005structural}.

\section{Numerical examples}
\label{ch:ex}

In this section, we demonstrate the capabilities of the proposed framework through several numerical examples. Since no physical units are used, the results may be interpreted under any consistent unit system. The crack radius is set to $1\%$ of the characteristic domain size in order for the topological-derivative-based ERR expression in Eq.~\eqref{eq:3DERR} to remain valid~\citep{alidoost2020energy}. Unless stated otherwise, the Young's moduli are taken as $E_s = 1.0$ and $E_v = 10^{-6}$ for  solid and void phases, respectively. The Poisson ratio for both materials is assumed to be $\nu = 0.3$. A value of $p = 8$ is used in the objective function Eq.~\eqref{eq:chJ}. The grid of compactly supported RBFs has the same spacing as the finite element mesh size (we use structured finite element discretizations). The support radius of the RBFs is set to twice the grid spacing.

\subsection{Triaxially loaded cube: shape optimization}
\label{sec:cubeElli}

\begin{figure}[!b]
  \centering
  \includegraphics{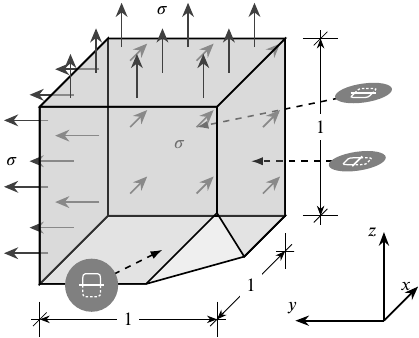}
  \caption{Reduced model exploiting the symmetry of the geometry and loading (triaxial tractions), showing one-eighth of the cube and one-eighth of the centered octahedral void, which appears as a tetrahedron in this reduced domain.}
  \label{fig:cube}
\end{figure}

We first consider a shape optimization problem. Consider a cube with an initial centered octahedron void subjected to triaxial tensile loading. 
Owing to symmetry in the geometry and loading, only one-eighth of the cube need be modeled, as illustrated in \Cref{fig:cube}, where the void appears as a tetrahedron.
 The cubic domain has side lengths $a = 1$ and is subjected to uniform tractions $\bar{\bs{t}}_i = \sigma \, \bs{e}_i, \, i = \{ x, y, z \} $ applied over its external boundary.

\begin{figure}[t]
  \centering
  \includegraphics{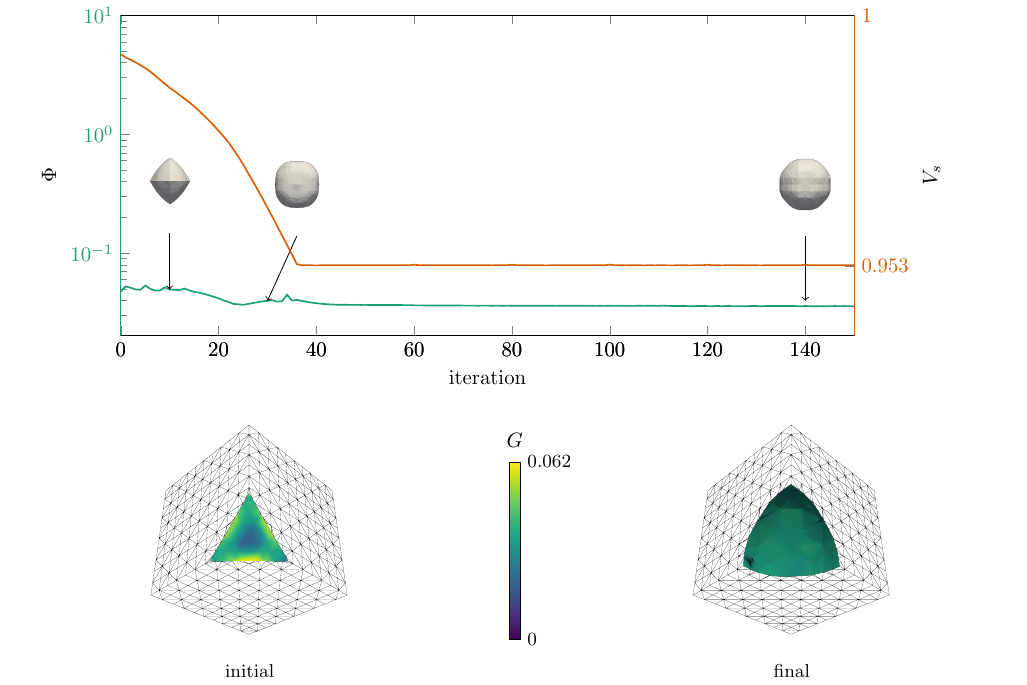}
  \caption{(top) Objective function $\Phi$ and normalized volume $V_s$ throughout the optimization process, with representative void configurations at iterations 10, 30, and 140. (bottom) ERR fields (nonzero only along the enriched free surface) for the initial and final designs; the final design converges to one-eighth of a spherical cavity. At convergence, the ERR is distributed uniformly over the free surface.}
  \label{fig:cube plot}
\end{figure}

The cubic domain is then discretized using $10 \times 10 \times 10 \times 6$ linear tetrahedral elements. \Cref{fig:cube plot} shows the discretization of the solid phase, which is the result of the interaction between the original mesh and the IGFEM geometric engine. A solid volume constraint $V_c = 96\%$ is imposed, defined relative to the initial design volume ($V_c = 95.3\%$ with respect to the complete cube). The maximum number of iterations in the optimization is set to 150, and no additional convergence criteria are introduced. The regularization of the LSF was initiated at iteration 60 and subsequently applied every 20 iterations to preserve the signed-distance property.

\Cref{fig:cube plot} shows the convergence history of both the objective function $\Phi$ and the solid volume $V_s$. For this optimization, the LSF is interpolated quadratically and regularized (re-initialized) every 20 iterations. The designs at iterations 10, 30, and 140 are also depicted to illustrate the morphological evolution of the design towards a spherical cavity, consistent with the expectation that a spherical void minimizes stress concentration---and hence the ERR---and in agreement with the findings of \citet{zhang2021on} for the analogous two-dimensional problem. Indeed, the resulting hole radii along the coordinate directions are nearly equal, $r_x = 0.443$, $r_y = 0.435$, and $r_z = 0.429$, confirming this near-spherical convergence.

Convergence is achieved after approximately 55 iterations, once the volume constraint is satisfied and the ERR distribution along the material interface becomes increasingly uniform. The corresponding ERR fields, nonzero only along the enriched free surface, are shown for both the initial and optimized designs; for the latter, $G = 0.0358$. Separately, we note that the periodic regularization of the LSF induces small perturbations in the volume, which in turn activate the volume constraint---an expected behavior, since the re-initialization step slightly modifies the design variables to restore the signed-distance property, introducing minor fluctuations in the constraint history.

 \begin{figure}[b]
  \begin{subfigure}[b]{0.48\textwidth}
    \centering
    \includegraphics[width=\linewidth]{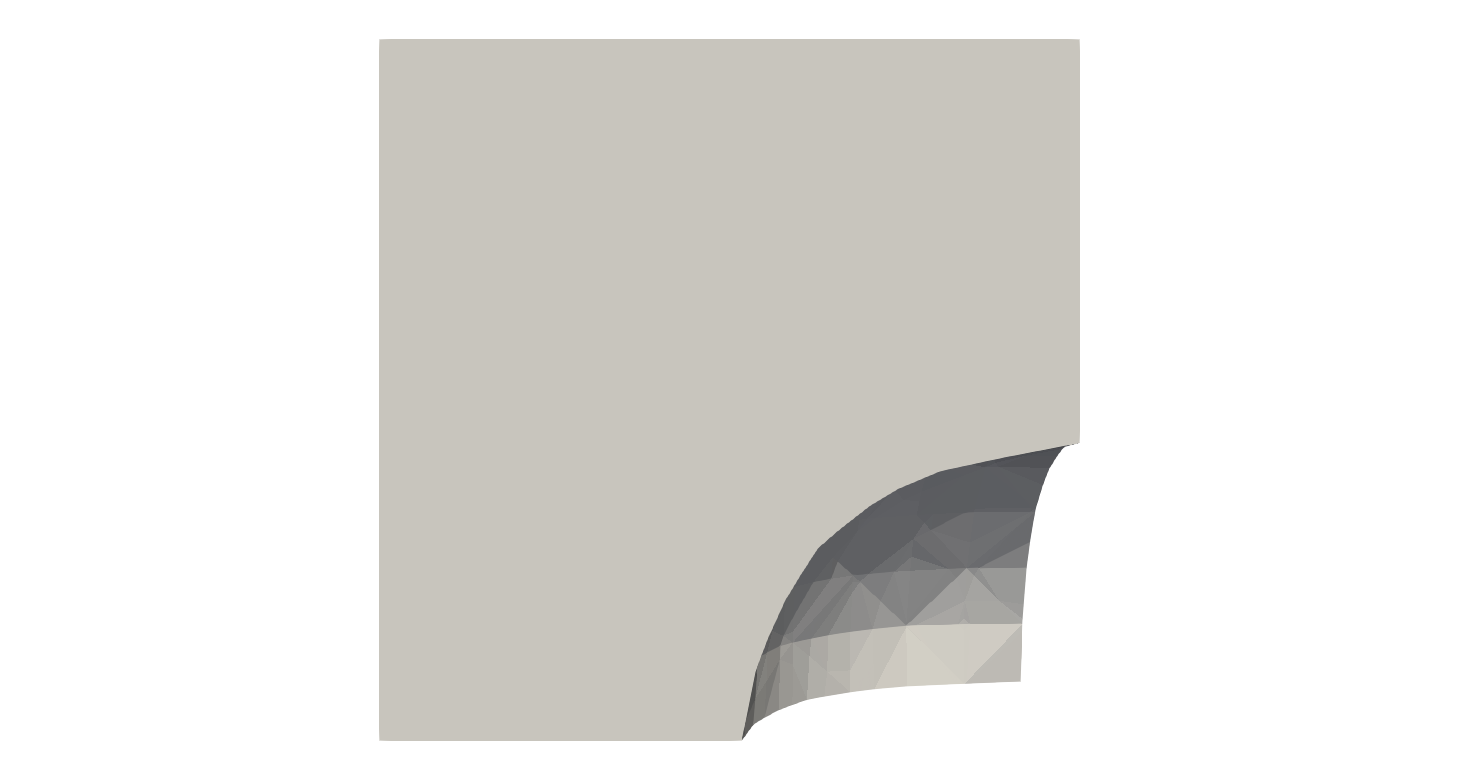}
    \caption{}
    \label{fig:Linear}
  \end{subfigure}
  \begin{subfigure}[b]{0.48\textwidth}
     \includegraphics[width=\linewidth]{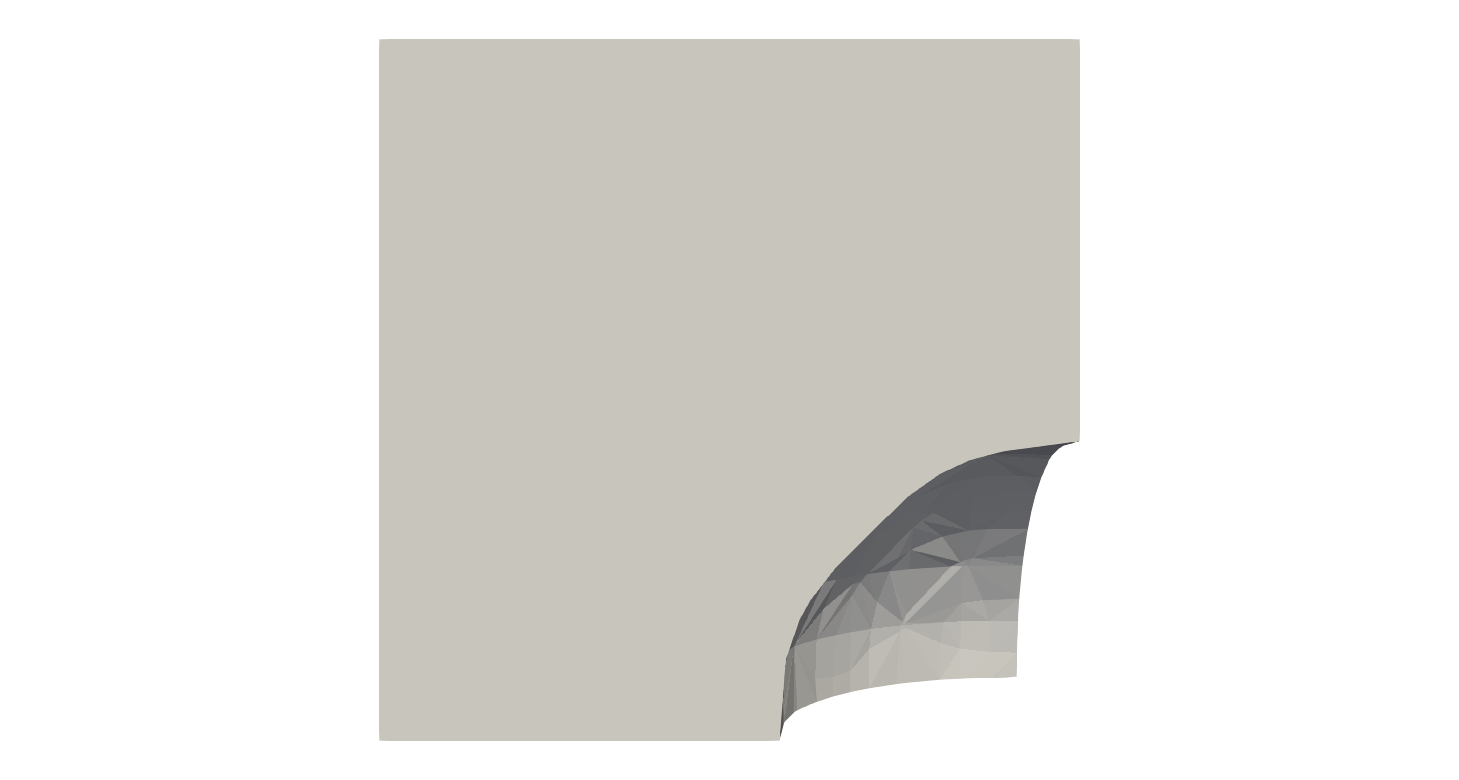}
    \caption{}
    \label{fig:NoReg}
  \end{subfigure}
  \caption{Side views of final optimized configurations for a linear interpolation of the LSF regularization (a), and a quadratic interpolation of the LSF without reinitialization (b).}
\label{fig:Comparison}
\end{figure}

To assess the role of the LSF interpolation and regularization, we solved the same problem with two additional configurations: (a) A linear LSF approximation with regularization, using the same re-initialization strategy (starting after 60 iterations and repeated every 20 iterations), and (b) A quadratic LSF approximation without regularization. The corresponding results are shown in \Cref{fig:Comparison}.
For the linear approximation, under the predefined mesh resolution, the optimizer fails to converge to a spherical cavity.
For the quadratic approximation without re-initialization, the behavior is different: the optimization initially captures a smooth spherical shape, but this quality is not preserved. As the signed-distance property is no longer enforced, geometric distortions gradually accumulate, ultimately leading to a final design that deviates from one-eight of a sphere.

\subsection{Triaxially loaded cube: topology optimization}

The previous example demonstrated that, for a fixed topology, the proposed methodology morphs the cavity's free surface towards a spherical shape with uniform ERR. This raises the question of whether such a solution remains optimal once topological changes are permitted. To investigate this, we revisit the problem and allow for the nucleation of new holes via the topological derivative criterion discussed in \Cref{ch:TD}. The geometry, loading conditions, material properties, finite element discretization, initial design, and optimization parameters are identical to those in the previous example.

\begin{figure}[h]
  \centering
  \includegraphics{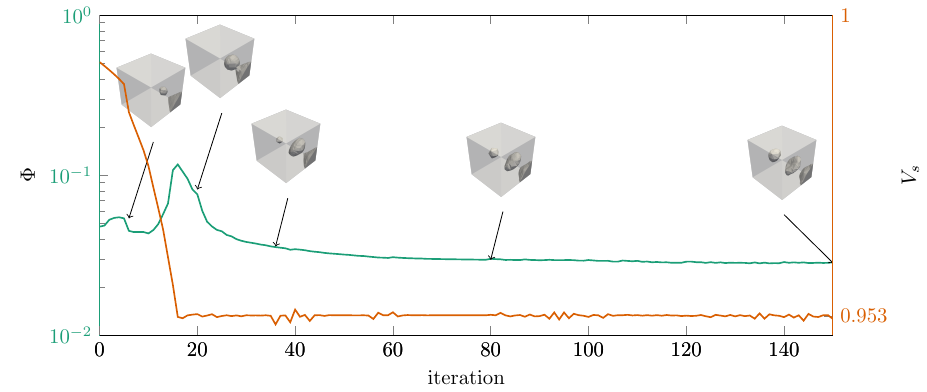}
  \caption{Objective function $\Phi$ and normalized solid volume $V_s$ throughout the topology optimization process. Representative designs at selected iterations illustrate the nucleation, merging, and subsequent separation of cavities during the optimization.}
  \label{fig:cube_TD}
\end{figure}

The optimization history is shown in \Cref{fig:cube_TD}, where we see that another cavity is nucleated after only six iterations. Initially, this cavity grows toward the original one and the two cavities temporarily merge, as shown at iteration 20. As the optimization progresses, a third cavity nucleates at iteration 36, after which the cavities gradually separate and evolve independently. The final design therefore consists of three disconnected cavities, each of which continues to undergo further shape refinement.

Compared with the shape optimization result of the previous section, the additional topological freedom reduces the objective function from 0.0358 to 0.0286, an improvement of approximately 20\%. This reduction is accompanied by a decrease in the local ERR values, whose range changes from $\left[0.0295, 0.0430\right]$ for the shape optimization to $\left[0.0079, 0.0407\right]$ when topological changes are allowed. This demonstrates that, although the spherical cavity is optimal under a fixed-topology assumption, enabling hole nucleation identifies a superior design consisting of multiple cavities. The example thus highlights the importance of combining boundary evolution with topological derivatives when minimizing energy release rates.

\subsection{Three-mode loading cube}

In this section, we demonstrate the capability of the proposed methodology to tailor the geometry of an internal cylindrical void so as to minimize the ERR when the domain is subjected to loading modes that mimic the classical opening (mode I), shearing (mode II), and tearing (mode III) modes of linear elastic fracture mechanics.

\begin{figure}[t]
  \centering
  \includegraphics{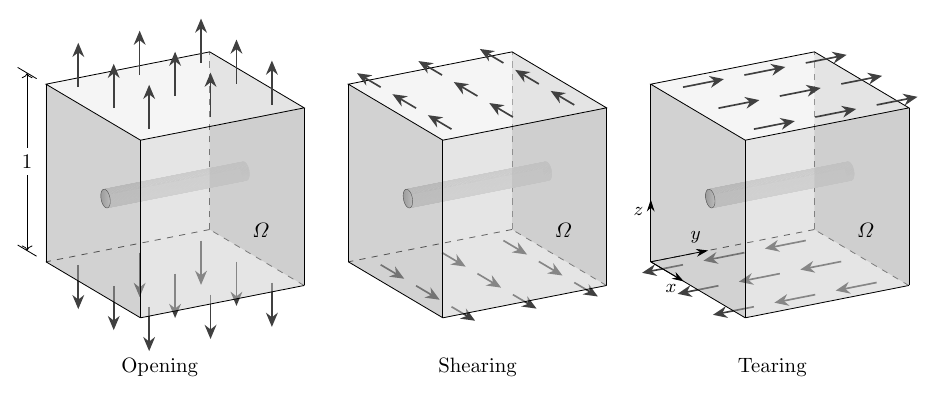}
  \caption{Unit cube with a cylindrical void as initial design subject to opening, shearing, and tearing loading modes.}
  \label{fig:three-mode_schematic}
\end{figure}

Consider the unit cube $\varOmega = [0,1]^3$ in \Cref{fig:three-mode_schematic}, subjected to tractions applied on the top and bottom faces. The initial design has a centered cylindrical hole of radius $r_v = 0.1$ connecting two opposite faces of the cubic domain, which is discretized over a $12 \times 12 \times 12$ Cartesian grid (totaling \num{10368} tetrahedra). In addition, an equality volume constraint is imposed such that the volume of the hole remains equal to $3\%$ of the total domain volume throughout the optimization process. No topological derivatives are used here to nucleate additional cavities in the solid.

\begin{figure}[t]
  \centering
  \includegraphics{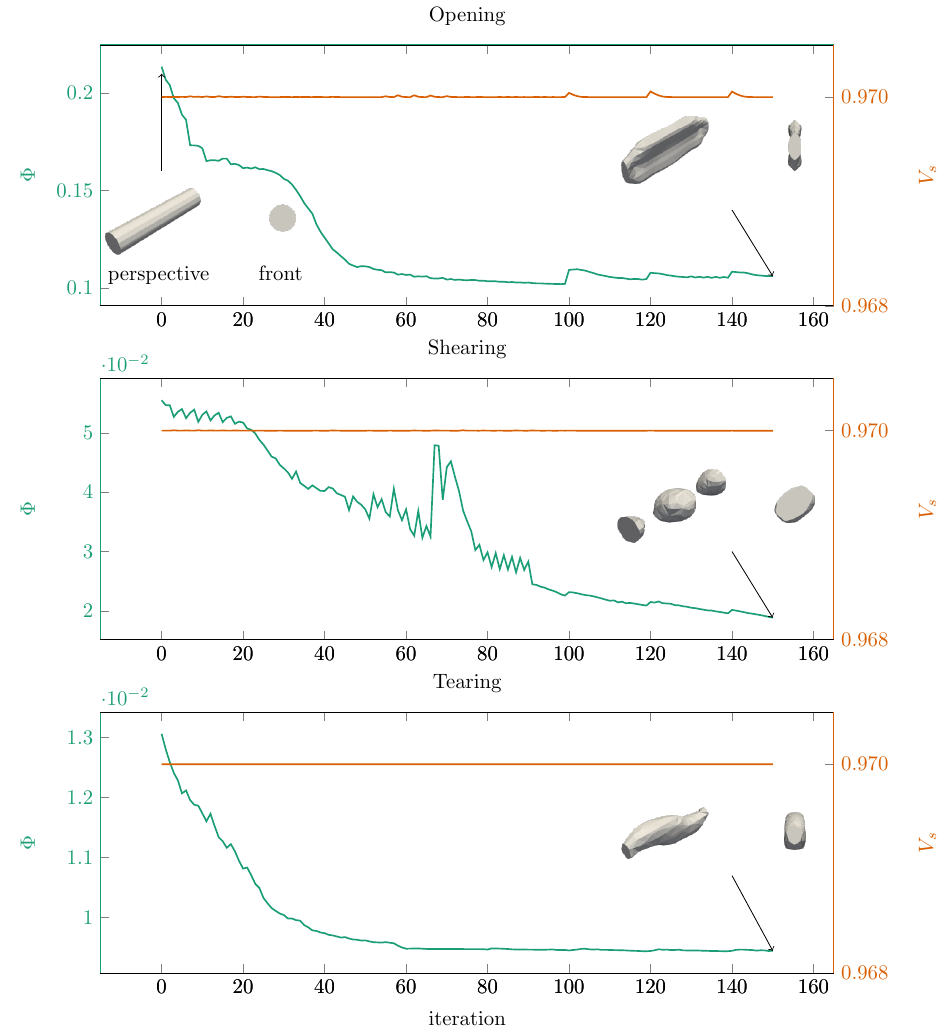}
  \caption{Objective function $\Phi$ and normalized solid volume $V_s$ as functions of iteration number for opening (top), shearing (middle) and tearing (bottom) loading modes. The cavity designs at the end of the optimization are also shown.}
  \label{fig:three-mode_histories}
\end{figure}

The optimization histories are shown in \Cref{fig:three-mode_histories}. We consider first the results for the opening mode, where it can be seen that the optimization transforms the initial cylindrical cavity into a narrow elongated cavity with a nearly uniform cross-section throughout its length. 
The resulting geometry increases the curvature, thereby alleviating stress concentrations around the cavity boundary. Minor reductions in cavity size near the cavity ends reflect three-dimensional effects, while the overall elongated morphology remains consistent throughout the domain. The resulting ERR field ranges from $G = \left[0, 0.25 \right]$ at iteration 1 to $G = \left[0, 0.14 \right] $ at iteration 150, where we notice a significant reduction of the peak ERR (about 44\% reduction).

Regarding the shearing mode, the optimized design rotates the cavity to align with the principal stress directions ($45^\circ$ in the $xz$ plane, see front view in the figure). The initially connected cavity also separates into three disconnected cavities: a dominant central cavity and two smaller side cavities. This redistribution reduces the ERR by concentrating most of the available void volume in the central region while relegating the remainder to the auxiliary cavities, suggesting that the optimization benefits jointly from stress-direction alignment and non-uniform volume redistribution.
The ERR field for this load mode ranges from $G = \left[ 0, 0.07 \right]$ at iteration 1 to $G = \left[ 0.0031, 0.0257 \right]$ at iteration 150, again showing a substantial reduction of the maximum ERR value on the cavity surface (about 63\%). The oscillation in the objective function observed between iterations 60 and 70 coincides with the separation of the initially connected cavity into three disconnected cavities. During this topological transition, thin ligaments temporarily connect the separating cavities, giving rise to localized stress concentrations and, consequently, elevated ERR values.

For the tearing mode, the optimization preserves a single connected void and introduces a gradual, wavy geometric modulation along the axis of the original cylindrical void. Comparison of the ERR distributions before and after optimization reveals a reduction in the dominant ERR hotspot, together with a more uniform distribution along the cavity surface. This design suggests that, under this loading mode, ERR minimization is achieved primarily through local geometric modulation rather than global reorientation or cavity separation. The ERR field range decreases from $G= \left[ 0, 0.0181 \right]$ to  $G= \left[0, 0.0133 \right]$, a reduction of approximately 27\%.

It is worth noting that the three loading modes exhibited different convergence characteristics throughout the optimization process. As a result, minor adjustments to the optimization (hyper)parameters were required to ensure stable and smooth convergence. Specifically, a move limit of $7.5 \times 10^{-5}$ was used for the opening mode, whereas a smaller value $5\times10^{-5}$ was adopted for the shearing and tearing modes. In addition, level-set regularization was activated only after 100 iterations for the opening and shearing cases, and after 60 iterations for the tearing case. Once activated, regularization was performed every 20 iterations for all three examples. These events caused slight spikes in the objective function value, more notably for the opening mode as shown in \Cref{fig:three-mode_histories}.

Overall, the results show strong agreement with established physical intuition regarding stress concentration and fracture behavior. The proposed method successfully captures the expected trends for each loading mode and their combinations, highlighting its potential for design applications involving fracture resistance.

\subsection{Immersed 3-D L-bracket}
\label{sec:L}

\begin{figure}
	\centering
	\includegraphics{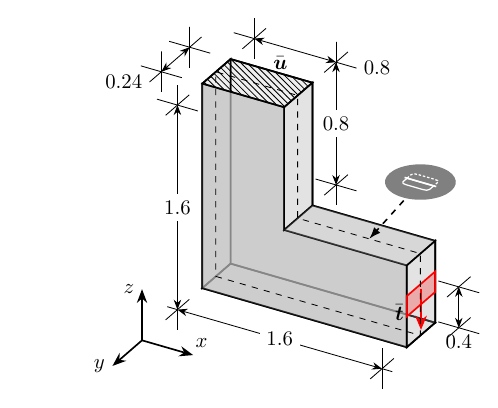}
	\caption{3-D L-bracket fixed on the top and subjected to a downward traction $\bar{\bs{t}}$ on the right end}
	\label{fig:L}
\end{figure}

The fourth and final example considers the classical L-bracket benchmark to assess the performance of the proposed framework on a larger-scale three-dimensional problem. The geometry, boundary conditions, and loading are shown in \Cref{fig:L}. The top surface of the bracket is fully clamped, while a downward traction, $\bar{\bs{t}} = - \bs{e}_z$, is applied at the other end of the bracket (see red shaded region in the figure). Owing to symmetry of the geometry, loading, and boundary conditions, only half of the structure is modeled, with symmetry enforced on the midplane parallel to $xz$.

Unlike the previous examples, the L-bracket is treated in a fully immersed setting: the entire solid is embedded within a cuboid, which is discretized using a $32 \times 32 \times 6$ Cartesian grid (\num{36864} linear tetrahedra). The cuboid domain spans $\left[0, 1.7\right] \times \left[0, 1.7\right] \times \left[0, 0.31875\right]$, resulting in a uniform element size $h = 1.7/32$. The L-bracket then occupies only a subset of this domain and is represented implicitly through the zeroth contour of the level-set function, leaving a layer of inactive elements around portions of the geometry. Finally, the optimization is initialized with a small hemispherical cavity located at the clamped boundary rather than with a completely solid design. This prevents the nucleation of voids directly at the supported boundary, which could otherwise lead to numerical instabilities.

To facilitate gradual nucleation of new holes, a continuation strategy is also adopted for the volume constraint. The optimization is initialized with a target volume fraction $V_c = 0.7$, which is subsequently reduced to $0.6$, $0.5$, and $0.4$ after 40, 60, and 80 iterations, respectively. This strategy allows newly nucleated holes sufficient time to evolve before further material removal. Furthermore, because substantial topological changes occur during the early optimization stages---due to both hole nucleation and volume-constraint updates---the first level-set regularization is postponed until iteration 80. Thereafter, regularization is performed every 20 iterations, consistent with the previous examples.

\begin{figure}[t]
  \centering
  \includegraphics{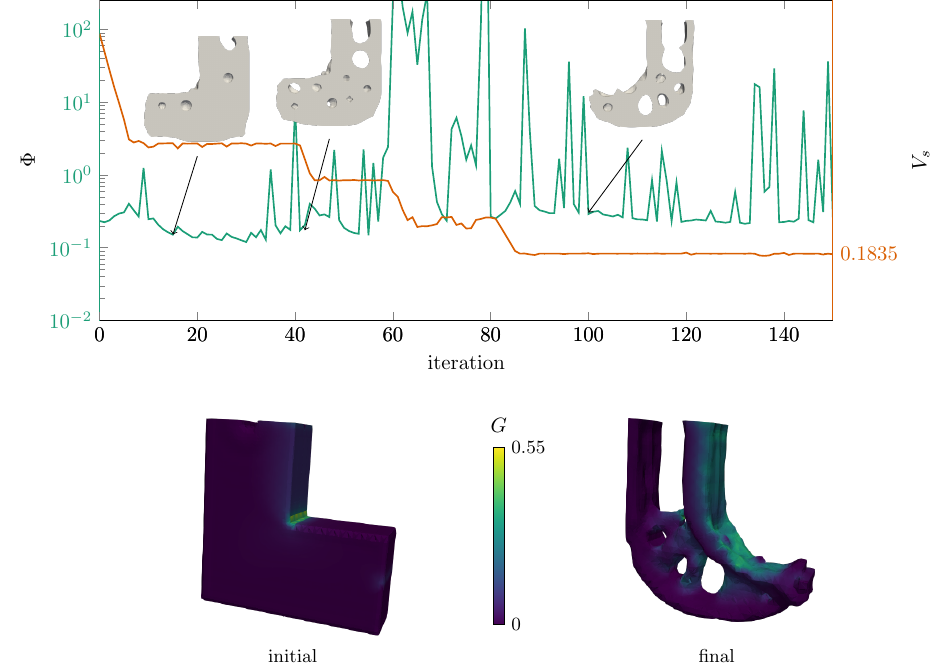}
  \caption{(top) Objective function $\Phi$ and normalized volume $V_s$ throughout the optimization process, with representative configurations (shown on the symmetry plane) at iterations 16, 42, and 100; (bottom) ERR fields throughout the free surface for the initial and final designs. The optimized design changes the shape of the initial sharp re-entrant corner to a smooth rounded surface, leading to a uniform distribution of ERRs.}
  \label{fig:Lbracket plot}
\end{figure}

The evolution of the design during the optimization process is illustrated in \Cref{fig:Lbracket plot}. Starting from a nearly solid design, new voids are introduced automatically through the topological derivative-based hole nucleation strategy. As the optimization progresses, additional holes emerge in regions identified as beneficial for improving fracture resistance, while existing boundaries undergo continuous shape refinement.

Several observations can be made from the optimization history. It is worth noting that in the first iterations of the optimization, the objective function has a lower value that at the end of the optimization. However, we notice that the volume constraint only converges after 80 iterations, so the lower value is due to the optimization rounding the re-entrant corner early in the optimization and the more bulky topology. This rounding continuous progressively, mitigating the stress concentration that would lead to crack initiation.

Wee see that the hole nucleation procedure enables substantial topology changes during the early stages of the optimization, after which the design evolves primarily through boundary motion and local shape refinement. The pronounced oscillations in the objective function are associated with the topological transitions, during which cavities merge or separate. These transient configurations introduce slender geometric features and small cut elements, leading to localized stress concentrations and temporarily elevated ERR values. Finally, the optimization converges to a well-defined layout characterized by smooth boundaries and a significantly modified material distribution, demonstrating the ability of the proposed methodology to combine topology and shape optimization within a fully immersed setting.

\section{Summary and Conclusions}
\label{ch:con}

In this work, we introduced a three-dimensional, fracture-driven topology optimization framework for reducing the likelihood of fracture in brittle solids. The framework combines a level-set description of geometry, discretized using compactly-supported radial basis functions, with the interface-enriched generalized finite element method (IGFEM) for analysis. Potential crack initiation sites are identified along material boundaries (at the location of enriched nodes), and their corresponding energy release rates are evaluated via topological derivatives following a single enriched finite element analysis of the uncracked domain. This avoids the need to explicitly model cracked geometries, which require specialized meshes for fracture assessment, resulting in a computationally efficient procedure suitable for optimization. Our procedure can be seen as a semi-analytical hybrid approach, in which stress intensity factors---and, subsequently, energy release rates---are computed accurately with the aid of weight functions calculated numerically, as described in \Cref{ssec:formulation}. To improve the accuracy of this computation, a three-dimensional stress recovery procedure was employed to alleviate the stress overestimation commonly observed in enriched finite element formulations. Local fracture indicators were aggregated using a $p$-mean function, allowing all potential crack locations to be incorporated into a single objective function. Analytical sensitivities were derived and verified through finite-difference analyses.

While \citet{zhang2021on} were the first to introduce quantities borrowed from linear elastic fracture mechanics into a topology optimization methodology---by means of topological derivatives---one may argue that their work allowed topological changes only through the merging of holes from an initial design. Their optimized designs were therefore largely dictated by the choice of initial design.
Beyond extending the original work of \citet{zhang2021on} to three dimensions, several methodological developments were introduced. These include a higher-order geometry representation through quadratic interpolation of the level-set function, a level-set regularization procedure to improve robustness throughout the optimization process, and a topological-derivative-based hole nucleation strategy that enables topology changes without prescribing the locations of new voids \textit{a priori}. Consequently, the proposed framework can perform both topology and shape optimization while maintaining an accurate description of evolving boundaries. The dual use of topological derivatives---both for hole nucleation and for fracture assessment during topology optimization---coupled with enriched finite element analysis, was termed $\partial^2 ( \mathrm{TO} ) $.

The numerical examples demonstrated the framework's ability to generate designs with significantly improved fracture resistance using fixed structured background meshes (also in a fully immersed setting). First, a cube subjected to tensile loading was used to verify the formulation and the sensitivity analysis. Starting from a design containing an octahedral cavity (a tetrahedron in the reduced domain, owing to symmetry), the optimization process evolved the cavity, ultimately converging to a geometry resembling the classical spherical-hole solution associated with a uniform energy release rate distribution along the free boundary. We further demonstrated that both the quadratic interpolation of the level-set function and the level-set regularization procedure are essential for obtaining the expected spherical cavity. Omitting either ingredient leads to geometric deterioration and prevents convergence to the anticipated optimal shape. Second, we then enabled hole nucleation in the same example and showed that a design with multiple cavities achieved an even lower energy release rate field. Third, we applied the framework to three benchmark problems that mimic modes I, II, and III of linear elastic fracture mechanics. These examples demonstrated the framework's ability to modify the geometry of an initial cylindrical cavity so as to mitigate fracture under these loading cases. Finally, we considered a larger-scale, fully immersed L-bracket example, in which the initial design consisted of a fully solid L-bracket implicitly described by the zeroth contour of the level-set function; topological derivatives were used not only to evaluate fracture resistance but also to nucleate new holes during the optimization process. The resulting design rounded the reentrant corner associated with the stress concentration, thereby significantly reducing the likelihood of fracture, while also highlighting the framework's capability to perform simultaneous topology and shape optimization on fixed structured background meshes. It is worth noting that, while similar designs can be obtained via stress-constrained topology optimization, in this work the design is instead obtained by evaluating energy release rates throughout the boundary (albeit derived from the stress field in the solid).

Our final conclusions are:
\begin{itemize}
\item Although the proposed topological-derivative-based hole nucleation strategy worked as expected, its performance remains dependent on user-defined parameters, such as the frequency of hole insertion and the level-set threshold used to identify candidate hole locations. These add further complexity to the (hyper)parameter set governing the optimization methodology. Future work should focus on developing more robust and automated nucleation criteria. Potential directions include adaptive hole-insertion strategies or alternative approaches based on reaction--diffusion equations coupled with the level-set method \citep{yamada2010topology}. The effect of other hyperparameters that define the methodology also needs further investigation, including optimizer parameters like the move limit used in MMA, the number of iterations between regularization of the level-set field, among others.

\item The methodology presented herein allows for a complete decoupling of the finite element discretization from both the geometric boundaries of the design and the grid of points defining the compactly supported radial basis functions. More research is needed to understand the effect of the (hyper)parameters involved in the discretization of the level-set field, including the grid spacing relative to the finite element size and the radius of the radial basis functions. Work in this direction is already underway in the context of electromagnetic topology optimization~\cite{Keller:2026}.

\item Oscillations in the optimization histories were observed. While some of these oscillations are physical---e.g., a slender, highly stressed strut connecting two holes before they merge---others are purely numerical; these emerge when interfaces cut tetrahedra in such a way that slender integration elements are created, thereby leading to an overestimation of the stress field. Although SIP improves the recovered stress field to a certain extent, by performing a non-local computation over a patch of elements that effectively blurs the field, we have observed situations where the stress remains overestimated. Eliminating these numerical artifacts requires modifying the underlying enriched finite element formulation to address the issue at its source.

\end{itemize}

To summarize, fracture resistance is highly dependent on the accurate representation of structural boundaries. By combining IGFEM with a level-set description of topology, the proposed framework enables the systematic design of brittle structures with improved fracture resistance through accurate geometry representation, efficient energy release rate evaluation, and seamless handling of evolving topologies on fixed background meshes. 

\appendix
\appendix

\crefname{section}{App.}{Appendices}
\Crefname{section}{App.}{Appendices}

\section{Devloo's Jacobian and its derivative}
\label[appendix]{Ap:devloo}

As discussed in \Cref{sec:boundcond}, the evaluation of the traction force vector in \Cref{eq:fe} requires integration over a surface element embedded in three-dimensional space. For a triangular surface element, the mapping is defined from a two-dimensional parent domain to the physical coordinates. Therefore, the corresponding Jacobian matrix is rectangular,

\begin{equation}
    \bs{J}_e
    =
    \bs{x}_e^\intercal
    \pdv{\bs{N}_e}{\boldsymbol{\xi}_e}
    =
    \begin{bmatrix}
        \pdv{x}{\xi} & \pdv{x}{\eta} \\
        \pdv{y}{\xi} & \pdv{y}{\eta} \\
        \pdv{z}{\xi} & \pdv{z}{\eta}
    \end{bmatrix},
    \label{eq:J}
\end{equation}
and hence it does not have a standard determinant or inverse. To obtain the surface measure, we follow the procedure of \citet{devloo1997pz}. Let
\begin{equation}
    \bs{a}=\pdv{\bs{x}}{\xi},
    \qquad
    \bs{b}=\pdv{\bs{x}}{\eta}.
\end{equation}
The tangent vectors are normalized as
\begin{equation}
    J_{\xi}=\norm{\bs{a}},
    \qquad
    J_{\eta}=\norm{\bs{b}},
\end{equation}
\begin{equation}
    \bs{V}_1=\frac{\bs{a}}{J_{\xi}},
    \qquad
    \widetilde{\bs{V}}_2=\frac{\bs{b}}{J_{\eta}}.
\end{equation}
The second orthonormal tangent direction is obtained by removing from $\widetilde{\bs{V}}_2$ its projection along $\bs{V}_1$,
\begin{equation}
    \bs{V}_2
    =
    \frac{
    \widetilde{\bs{V}}_2
    -
    \big(\widetilde{\bs{V}}_2\cdot\bs{V}_1\big)\bs{V}_1
    }
    {
    \left\|
    \widetilde{\bs{V}}_2
    -
    \big(\widetilde{\bs{V}}_2\cdot\bs{V}_1\big)\bs{V}_1
    \right\|
    }.
\end{equation}
With the above definitions, the rectangular Jacobian matrix in \Cref{eq:J} can be transformed into the equivalent Devloo Jacobian matrix
\begin{equation}
    \bs{J}_e
    =
    \begin{bmatrix}
        J_{\xi} &
        \big(\bs{V}_1\cdot\widetilde{\bs{V}}_2\big)J_{\eta}
        \\
        0 &
        \big(\bs{V}_2\cdot\widetilde{\bs{V}}_2\big)J_{\eta}
    \end{bmatrix}.
\end{equation}
The corresponding integration Jacobian is obtained from the determinant of the Devloo Jacobian matrix as
\begin{equation}
    j_e
    =
    J_{\xi}J_{\eta}
    \big(\bs{V}_2\cdot\widetilde{\bs{V}}_2\big).
    \label{eq:devloo_j}
\end{equation}

\section{Weight function evaluation}
\label[appendix]{Ap:weight}

\begin{figure}[!b]
	\centering
	\begin{subfigure}[b]{0.3\textwidth}
		\includegraphics{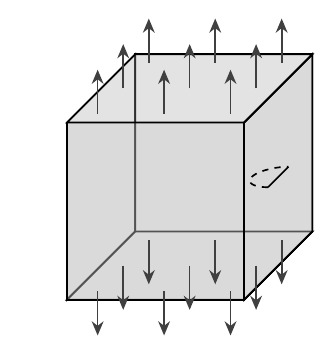}
		\caption{Mode I loading}
	\end{subfigure}
    \begin{subfigure}[b]{0.3\textwidth}
        \includegraphics{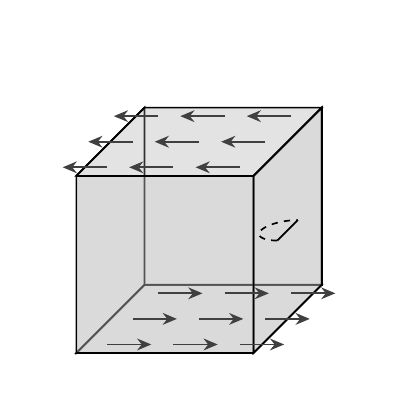}
	    \caption{Mode II loading}
    \end{subfigure}
    \begin{subfigure}[b]{0.3\textwidth}
        \includegraphics{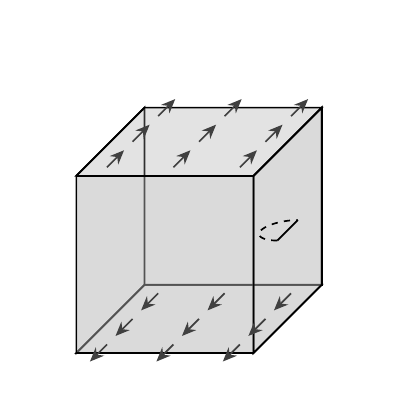}
    	\caption{Mode III loading}
    \end{subfigure}
    \caption{Test structure under mode-I, II, III loading}
    \label{fig:WFtest}
\end{figure}

\begin{figure}[t]
	\centering
	\begin{minipage}[t]{0.45\textwidth}
		\centering
		\includegraphics[width=\textwidth]{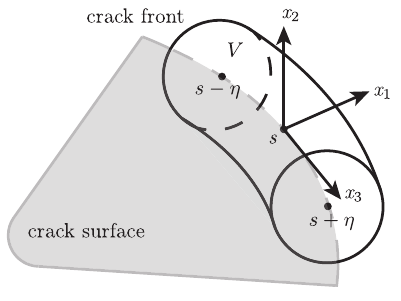}
		\caption{Integration domain at the crack front}
		\label{fig:domain}
	\end{minipage}
	\begin{minipage}[t]{0.45\textwidth}
		\centering
		\includegraphics[width=0.7\textwidth]{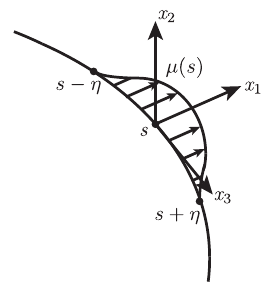}
		\caption{Virtual crack advance function $\mu$}
		\label{fig:mu}
	\end{minipage}
\end{figure}

To construct the weight function matrix, a cuboidal specimen containing a penny-shaped crack is analyzed under three independent loading conditions corresponding to Modes I, II, and III, as illustrated in \Cref{fig:WFtest}. For each loading case, the corresponding stress intensity factors are subsequently recovered from the interaction energy release rate. An uncracked counterpart with identical geometry and boundary conditions is analyzed to obtain the corresponding local stress field. The stress intensity factors from the cracked analyses and the stress fields from the uncracked analyses are subsequently used to determine the weight function matrix. The local interaction energy release rate, denoted as $M_1$, is evaluated using the interaction integral, defined as \citep{nakamura1989antisymmetrical, nagai2013stress, zhang2019stable}
\begin{equation}
\label{eq:M1}
	M_1 = \frac{\int_V \bigl(\sigma_{ij}^{(1)}\pdv{u_j^{(2)}}{x_1}+\sigma_{ij}^{(2)}\pdv{u_j^{(1)}}{x_1}-\sigma_{j1}^{(1)}\epsilon_{j1}^{(2)}\delta_{1i}\bigr) \pdv{q_1}{x_i} \, \dd{V}}{\int_{s-\eta}^{s+\eta} \mu(s) \, \dd{s}},
\end{equation}
where the integration domain $V(s)$ associated with the crack front position $s$ is illustrated in \Cref{fig:domain}. The domain has a length of $2\eta$ and a radius $r$, and is bounded by the circular surfaces located at $s-\eta$ and $s+\eta$. Here, $\sigma$, $\epsilon$, and $u$ denote the stress, strain, and displacement fields, respectively; $q_1$ is a weighting function that is continuous within $V(s)$ and vanishes outside the integration domain; $\mu$ denotes the virtual crack advance function at the crack front position $s$, as illustrated in \Cref{fig:mu}; and superscripts $(1)$ and $(2)$ refer to the actual and auxiliary fields, respectively.

The local interaction energy release rate can alternatively be expressed in terms of the stress intensity factors as
\begin{equation}
	M_1 = \frac{1-\nu^2}{E} \left( 2\SIFmodeI^{(1)}\SIFmodeI^{(2)} + 2\SIFmodeII^{(1)}\SIFmodeII^{(2)} \right) + \frac{1+\nu}{E}\left( 2\SIFmodeIII^{(1)}\SIFmodeIII^{(2)} \right).
\end{equation}
Equating \Cref{eq:M1} with the analytical expression above establishes a direct relation between the local interaction energy release rate and the stress intensity factors. By selecting the auxiliary fields as $(a)$ $[\SIFmodeI^{(2)}=1,\;\SIFmodeII^{(2)}=0,\;\SIFmodeIII^{(2)}=0]$, $(b)$ $[\SIFmodeI^{(2)}=0,\;\SIFmodeII^{(2)}=1,\;\SIFmodeIII^{(2)}=0]$, and $(c)$ $[\SIFmodeI^{(2)}=0,\;\SIFmodeII^{(2)}=0,\;\SIFmodeIII^{(2)}=1]$, the Mode I, II, and III stress intensity factors are obtained from
\begin{equation}
  \begin{cases}
    2\frac{1-\nu^2}{E} \SIFmodeI^{(1)} = J_a, \\[3pt]
    2\frac{1-\nu^2}{E} \SIFmodeII^{(1)} = J_b, \\[3pt]
    2\frac{1+\nu}{E} \SIFmodeIII^{(1)} = J_c.
  \end{cases}
\end{equation}

For each loading case, the interaction integral yields the corresponding stress intensity factors, while the uncracked analysis provides the local stress field. To eliminate the dependence of the stress intensity factors on the crack length $\epsilon$, the normalized stress intensity factors are defined as
\begin{equation}
\overline{\bs{K}} = \frac{\bs{K}}{\sqrt{\pi \epsilon}}
\end{equation}
Following the weight function formulation \citep{alidoost2020energy}, the stress intensity factors are assumed to be linearly related to the local stress field according to
\begin{equation}
\frac{\bs{K}}{\sqrt{\pi\epsilon}}=\bs{H}\,\boldsymbol\sigma,
\end{equation}
where $\epsilon$ denotes the crack length. Stacking the three independent loading cases gives
{\renewcommand{\arraystretch}{1.4}
\begin{equation}
	\frac{1}{\sqrt{\pi \epsilon}}
	\begin{bmatrix}
		\SIFmodeI^{(i)} & \SIFmodeI^{(ii)} & \SIFmodeI^{(iii)} \\
		\SIFmodeII^{(i)} & \SIFmodeII^{(ii)} & \SIFmodeII^{(iii)} \\
		\SIFmodeIII^{(i)} & \SIFmodeIII^{(ii)} & \SIFmodeIII^{(iii)}
	\end{bmatrix}
    =
    \bs{H}
    \begin{bmatrix}
    	\sigma_{\theta \theta}^{(i)} & \sigma_{\theta \theta}^{(ii)} & \sigma_{\theta \theta}^{(iii)} \\
    	\sigma_{r \theta}^{(i)} & \sigma_{r \theta}^{(ii)} & \sigma_{r \theta}^{(iii)} \\
    	\sigma_{z \theta}^{(i)} & \sigma_{z \theta}^{(ii)} & \sigma_{z \theta}^{(iii)} \\
    \end{bmatrix}.
\end{equation}}
solving the resulting linear system yields the weight function matrix $\bs{H}$.
\section{Energy release rate computation}
\label[appendix]{Ap:ERR_comp}

The ERR expression given in \Cref{eq:befERRfinal} is formulated in a local cylindrical coordinate system attached to the crack. In the present work, the crack is assumed to nucleate locally normal to the boundary, while its opening direction is taken to coincide with the direction of the maximum hoop stress. These assumptions provide a unique definition of the local crack coordinate system at every evaluation point along the boundary. However, determining the maximum hoop stress direction directly would require solving a principal stress eigenvalue problem, or equivalently the associated Cardano polynomial, at every evaluation point. Such a procedure would significantly increase the computational cost and substantially complicate the derivation of the corresponding sensitivity expressions.

Therefore, a different strategy is adopted in the present work. Since the crack is assumed to nucleate normal to the boundary, the radial direction $\bs{e}_r$ is uniquely determined by the outward boundary normal. The remaining tangential directions, however, are not unique. As an intermediate reference configuration, they are first chosen to coincide with the local $xz$-plane. Accordingly, the stress tensor is transformed from the global coordinate system into this local reference frame as

\begin{equation}
    \begin{bmatrix}
        \sigma_{\theta \theta} \\
        \sigma_{r \theta} \\
        \sigma_{z \theta}
    \end{bmatrix}
    =
    \underbrace{\begin{bmatrix}
        0 & 1 & 0 \\
        1 & 0 & 0 \\
        0 & 0 & 1
    \end{bmatrix}}_\text{$\bs{M}_1$}
    \bs{R}
    \underbrace{\begin{bmatrix}
        	\sigma_{xx} & \sigma_{xy} & \sigma_{xz} \\
        	\sigma_{yx} & \sigma_{yy} & \sigma_{yz} \\
        	\sigma_{zx} & \sigma_{zy} & \sigma_{zz}
        \end{bmatrix}}_\text{$\bs{\sigma}$}
    \bs{R}^\intercal
    \underbrace{\begin{bmatrix}
        	0 \\
        	1 \\
        	0
        \end{bmatrix},}_\text{$\bs{M}_2$}
    \label{eq:sigmaTrans}
\end{equation}
where the rotation matrix $\bs{R}$ is given in \Cref{R}. Substituting the transformed stresses into~\Cref{eq:ERRfinal} yields
\begin{equation}
	G = \frac{\pi \epsilon}{2 \mu \bar{E}} \bs{M}_2^\intercal \bs{R} \, \bs{\sigma}^\intercal \bs{R}^\intercal \bs{M}_1^\intercal {\bs{H}'}^\intercal \bs{H}' \bs{M}_1 \bs{R} \, \bs{\sigma} \bs{R}^\intercal \bs{M}_2.
	\label{eq:ERR_m}
\end{equation}
\Cref{eq:sigmaTrans} transforms the stress tensor into the local coordinate system whose normal is aligned with the crack normal, while while the tangential basis vectors are associated with the auxiliary $xz$-plane. The remaining step consists of an additional in-plane transformation that aligns the local coordinate system with the direction of maximum hoop stress. Accordingly, \Cref{eq:sigmaTrans} becomes
\begin{equation}
    \begin{bmatrix}
        \sigma_{\theta \theta} \\
        \sigma_{r \theta} \\
        \sigma_{z \theta}
    \end{bmatrix}
    =
    \underbrace{\begin{bmatrix}
        0 & 1 & 0 \\
        1 & 0 & 0 \\
        0 & 0 & 1
    \end{bmatrix}}_\text{$\bs{M}_1$}
    \bs{R}_t
    \bs{R}
    \underbrace{\begin{bmatrix}
        	\sigma_{xx} & \sigma_{xy} & \sigma_{xz} \\
        	\sigma_{yx} & \sigma_{yy} & \sigma_{yz} \\
        	\sigma_{zx} & \sigma_{zy} & \sigma_{zz}
        \end{bmatrix}}_\text{$\bs{\sigma}$}
    \bs{R}^\intercal
    \bs{R}_t^\intercal
    \underbrace{\begin{bmatrix}
        	0 \\
        	1 \\
        	0
        \end{bmatrix}}_\text{$\bs{M}_2$},
    \label{eq:principalTrans}
\end{equation}
where the rotation matrix $\bs{R}_t$ defines the additional local transformation associated with the maximum hoop stress direction and is given in \Cref{R}. Including this additional transformation, the ERR expression in~\eqref{eq:ERR_m} becomes
\begin{equation}
	G = \frac{\pi \epsilon}{2 \mu \bar{E}} \bs{M}_2^\intercal \bs{R}_t \bs{R} \bs{\sigma}^\intercal \bs{R}^\intercal \bs{R}_t^\intercal \bs{M}_1^\intercal \bs{H'}^\intercal \bs{H'} \bs{M}_1 \bs{R}_t \bs{R} \, \bs{\sigma} \bs{R}^\intercal \bs{R}_t^\intercal \bs{M}_2,
	\label{eq:ERR_principal}
\end{equation}
which can be rewritten in the compact form
\begin{equation}
    G = \frac{\pi \epsilon}{2 \mu \bar{E}} \bs{Q}_R^\intercal \boldsymbol{\sigma}^\intercal \bs{F}\boldsymbol{\sigma} \bs{Q}_R,
\end{equation}
where $\bs{Q}_R = \bs{R}^\intercal \bs{R}_t^\intercal \bs{M}_2$ and $\bs{F} = \bs{R}^\intercal \bs{R}_t^\intercal \bs{M}_1^\intercal \bs{H'}^\intercal \bs{H'} \bs{M}_1 \bs{R}_t \bs{R} $

Finally, following the classical hypothesis for brittle fracture propagation~\citep{erdogan1963crack}, cracks are assumed to propagate perpendicular to the direction of maximum tensile stress. Consequently, only crack configurations associated with positive mode-I stress intensity factors, $ \SIFmodeI > 0 $, are considered in the ERR evaluation and subsequent aggregation in~\eqref{eq:chJ}.

\section{Coordinate Transformation Matrix}
 \label[appendix]{R}

The local crack coordinate system is constructed based on a nodal surface normal vector. Since the computational domain is discretized with tetrahedral elements, the external boundary is composed of triangular surface facets corresponding to boundary faces of tetrahedral elements. Because a surface node is generally shared by several boundary facets, the nodal normal vector is obtained by averaging the outward unit normals of all connected surface facets. Accordingly,
\begin{equation}
    \hat{\bs{n}}
    =
    \frac{1}{N_s}
    \sum_{i=1}^{N_s}
    \bs{n}_i, \quad \bs{n} = \frac{\hat{\bs{n}}}{{\left\lVert \bs{\hat{\bs{n}}}\right\rVert},}
    \label{eq:nodal_normal}
\end{equation}
where $N_s$ denotes the number of boundary surface facets sharing the node, and $\bs{n}_i=(n_{ix},n_{iy},n_{iz})$ is the outward normal vector of the $i$th triangular surface facet.
The elemental surface normal is evaluated from the cross product of two edge vectors of the triangular facet as
\begin{equation}
    \bs{n}_i
    =
    m_i \left(\bs{d}_{01}\times\bs{d}_{02}\right)
\label{eq:nd}
\end{equation}
where $\bs{d}_{01}$ and $\bs{d}_{02}$ are two edge vectors originating from the same vertex of the triangular surface facet, where $m_i \in \{-1, 1\}$ is selected such that $\bs{n}_i$ is oriented outward.

\subsection{Crack opening parallel to $xz$-plane}
To define the local crack coordinate system for a crack whose opening direction is assumed to lie in the $xz$-plane, three orthonormal vectors are required. The starting point is the global Cartesian coordinate system $(x,y,z)$ with basis vectors $(\bs{e}_x,\bs{e}_y,\bs{e}_z)$. The local coordinate system $(x',y',z')$ is then constructed from the surface normal vector and the assumed crack opening direction. The first local direction, denoted by $\bs{k}$, is defined as the inward surface normal, 
\begin{equation}
    \bs{k} = -\bs{n} = (-n_x,-n_y,-n_z),
\end{equation}
which defines the $x'$ direction corresponding to the deepest crack-front direction. The second direction, denoted by $\bs{m}$, defines the local $z'$ axis. This vector is required to satisfy two conditions. First, it must be perpendicular to the surface normal vector $\bs{n}$. Second, since the crack opening direction is assumed to lie in the global $xz$-plane, $\bs{m}$ must also be perpendicular to the global $y$-direction, i.e. to the vector $\bs{e}_y = (0,1,0)$. Enforcing these orthogonality conditions together with the unit-length constraint yields
\begin{equation}
	\begin{cases}
		\bs{m} \cdot \bs{n} = 0, \\
		\bs{m} \cdot (0, 1 , 0) = 0, \\
		|\bs{m}| = 1,
	\end{cases}
	\label{eq:eq_nz}
\end{equation}
Solving~\eqref{eq:eq_nz} gives
\begin{equation}
	\bs{m} = \begin{pmatrix}
	 -\frac{n_z}{\sqrt{n_x^2+n_z^2}} , \; 0 , \;  \frac{n_x}{\sqrt{n_x^2+n_z^2}} \end{pmatrix}.
	\label{eq:nz}
\end{equation}
The third local direction, $\bs{l}$, associated with the $y'$ axis, is then obtained from the cross product between the first two directions,
\begin{equation}
    \bs{l} = -\bs{n} \times \bs{m},
\end{equation}
which leads to
\begin{equation}
    \bs{l} = \frac{1}{\sqrt{n_x^2+n_z^2} \sqrt{n_x^2+n_y^2+n_z^2}} \begin{pmatrix} -n_x n_y , \;  n_x^2+n_z^2 , \; -n_y n_z  \end{pmatrix}.
    \label{eq:ny}
\end{equation}
Once the three orthonormal directions are defined, the transformation between the global and local coordinate systems can be written as
{\renewcommand{\arraystretch}{2}
\begin{equation}
    \bs{R}
    \begin{bmatrix}
	-n_x & \frac{-n_x n_y}{\sqrt{n_x^2+n_z^2} \sqrt{n_x^2+n_y^2+n_z^2}} & -\frac{n_z}{\sqrt{n_x^2+n_z^2}} \\
	-n_y & \frac{n_x^2+n_z^2}{\sqrt{n_x^2+n_z^2} \sqrt{n_x^2+n_y^2+n_z^2}} & 0 \\
	-n_z & \frac{-n_y n_z}{\sqrt{n_x^2+n_z^2} \sqrt{n_x^2+n_y^2+n_z^2}} & \frac{n_x}{\sqrt{n_x^2+n_z^2}}
    \end{bmatrix}
    =
    \begin{bmatrix}
        1 & 0 & 0 \\
        0 & 1 & 0 \\
        0 & 0 & 1
    \end{bmatrix}.
    \label{eq:eq_R1}
\end{equation}
}
Solving~\Cref{eq:eq_R1} yields the transformation matrix
{\renewcommand{\arraystretch}{2}
\begin{equation}
    \bs{R} = 
    \begin{bmatrix}
	\frac{-n_x}{n_x^2+n_y^2+n_z^2} & \frac{-n_y}{n_x^2+n_y^2+n_z^2} & \frac{-n_z}{n_x^2+n_y^2+n_z^2} \\
	\frac{-n_x n_y}{\sqrt{n_x^2+n_z^2} \sqrt{n_x^2+n_y^2+n_z^2}} & \frac{\sqrt{n_x^2+n_z^2}}{\sqrt{n_x^2+n_y^2+n_z^2}} & \frac{-n_y n_z}{\sqrt{n_x^2+n_z^2} \sqrt{n_x^2+n_y^2+n_z^2}} \\
	-\frac{n_z}{\sqrt{n_x^2+n_z^2}} & 0 & \frac{n_x}{\sqrt{n_x^2+n_z^2}}
    \end{bmatrix}.
    \label{eq:R1}
\end{equation}
}

\begin{figure}[t]
    \centering
\includegraphics{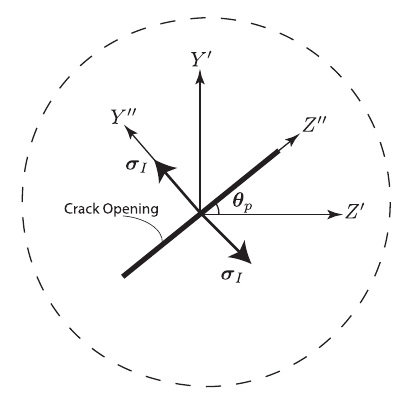}
    \caption{Determination of the crack opening direction from the surface first principal stress. The local coordinate system $(X',Y',Z')$ is first constructed using the surface normal and the auxiliary crack-opening assumption. A subsequent rotation by $\theta_p$ within the $Y'Z'$ plane aligns the transformed coordinate system $(X'',Y'',Z'')$ with the principal stress direction $\sigma_I$. The $X''$ axis remains coincident with $X'$.}
    \label{fig:crack_principal}
\end{figure}

\subsection{Crack opening determined by surface principal stress}

After transforming the stress tensor into the local coordinate system associated with the crack plane and the auxiliary $xz$-plane, the normal to the crack plane is uniquely determined, whereas the orientation of the local tangential directions remains arbitrary. A final transformation is therefore introduced to align the crack opening direction with the direction of maximum hoop stress. As illustrated in \Cref{fig:crack_principal}, this transformation consists of a rotation within the local $y'z'$-plane from the auxiliary coordinate system $(x',y',z')$ to the final crack coordinate system $(x'',y'',z'')$. Since the rotation is confined to the tangent plane, the $x'$ axis remains unchanged. Accordingly, the stress transformation matrix $\bs{R}_t$ takes the form
\begin{equation}
    \bs{R}_t
    =
    \begin{bmatrix}
        1 & 0 & 0 \\
        0 & \cos(\theta_p) & -\sin(\theta_p) \\
        0 & \sin(\theta_p) & \cos(\theta_p)
    \end{bmatrix},
\end{equation}
where $\theta_p$ denotes the angle between the direction of the surface first principal stress and the $y'$ axis. The angle $\theta_p$ is evaluated as~\cite{sun2009mat}
\begin{equation}
    \theta_p =
    \frac{1}{2}
    \arctan
    \Big(
    -\frac{2 \sigma_{y'z'}}
    {\sigma_{z'z'}- \sigma_{y'y'}}
    \Big),
    \label{eq:principal_angle}
\end{equation}
where $\sigma_{y'z'}$, $\sigma_{y'y'}$, and $\sigma_{z'z'}$ denote the stress components expressed in the local coordinate system. The orientation associated with the maximum principal stress is selected for the crack opening direction.

\section{3D Stress Recovery Technique}
\label[appendix]{Ap:matrices}

The interpolation matrices and differential operator of the stress recovery technique in \Cref{eq:stress} are:

\begin{equation}
	\bs{E}_{\sigma} =
	\begin{bmatrix}
		\bs{e}_{\sigma} & & & & & \\
		& \bs{e}_{\sigma} & & & & \\
		& & \bs{e}_{\sigma} & & & \\
		& & & \bs{e}_{\sigma} & & \\
		& & & & \bs{e}_{\sigma} & \\
		& & & & & \bs{e}_{\sigma} \\
	\end{bmatrix},
\end{equation}

\begin{equation}
	\bs{e}_{\sigma} =
	\begin{bmatrix}
		1 & x & y & z & xy & yz & zx & x^2 & y^2 & z^2
	\end{bmatrix},
\end{equation}

\begin{equation}
	\partial_{\sigma}=
	\begin{bmatrix}
		\pdv{x} & 0 & 0 & \pdv{y} & 0 & \pdv{z} \\
		0 & \pdv{y} & 0 & \pdv{x} & \pdv{z} & 0 \\
		0 & 0 & \pdv{z} & 0 & \pdv{y} & \pdv{x} \\
	\end{bmatrix},
\end{equation}

\setcounter{MaxMatrixCols}{20}
\begin{equation}
	\bs{E}_{\zeta} = 
	\begin{bmatrix}
		1 & x & y & z & 0 & 0 & 0 & 0 & 0 & 0 & 0 & 0\\
		0 & 0 & 0 & 0 & 1 & x & y & z & 0 & 0 & 0 & 0\\
		0 & 0 & 0 & 0 & 0 & 0 & 0 & 0 & 1 & x & y & z\\
	\end{bmatrix},
\end{equation}

\begin{equation}
	\bar{\bs{E}}_{\sigma}^\intercal=
	\begin{bmatrix}
		1 & 0 & 0 & 0 & 0 & 0 \\
		0 & 1 & 0 & 0 & 0 & 0 \\
		0 & 0 & 1 & 0 & 0 & 0 \\
		0 & 0 & 0 & 1 & 0 & 0 \\
		0 & 0 & 0 & 0 & 1 & 0 \\
		0 & 0 & 0 & 0 & 0 & 1 \\
		x & 0 & 0 & -y & 0 & 0 \\
		y & 0 & 0 & 0 & 0 & 0 \\
		z & 0 & 0 & 0 & 0 & 0 \\
		2xy & 0 & 0 & -y^2 & 0 & 0 \\
		yz & 0 & 0 & 0 & 0 & 0 \\
        2xz & 0 & 0 & 0 & 0 & -z^2 \\
		x^2 & 0 & 0 & -xy & yz & -xz \\
		0 & 2x & 0 & 0 & 0 & 0 \\
		0 & y & 0 & 0 & -z & 0 \\
		0 & 2z & 0 & 0 & 0 & 0 \\
		0 & 2xy & 0 & -x^2 & 0 & 0 \\
		0 & 2yz & 0 & 0 & -z^2 & 0 \\
		0 & xz & 0 & 0 & 0 & 0 \\
		0 & y^2 & 0 & -xy & -yz & xz \\
		0 & 0 & 2x & 0 & 0 & 0 \\
		0 & 0 & 2y & 0 & 0 & 0 \\
		0 & 0 & z & 0 & -y & 0 \\
		0 & 0 & xy & 0 & 0 & 0 \\
		0 & 0 & 2yz & 0 & -y^2 & 0 \\
		0 & 0 & 2xz & 0 & 0 & -x^2 \\
		0 & 0 & z^2 & xy & -yz & -xz \\
		0 & y & 0 & -x & 0 & 0 \\
		0 & 0 & 0 & -z & 0 & 0 \\
		0 & 0 & 0 & 2yz & 0 & -z^2 \\
		0 & 0 & 0 & 2xz & -z^2 & 0 \\
		0 & x^2 & 0 & 0 & 0 & 0 \\
		y^2 & 0 & 0 & 0 & 0 & 0 \\
		0 & 0 & 0 & -z^2 & 0 & 0 \\
		0 & 0 & 0 & 0 & -x & 0 \\
		0 & 0 & 0 & 0 & 2xy & -x^2 \\
		0 & 0 & 0 & -x^2 & 2xz & 0 \\
		0 & 0 & 0 & 0 & -x^2 & 0 \\
		0 & 0 & y^2 & 0 & 0 & 0 \\
		0 & z^2 & 0 & 0 & 0 & 0 \\
		0 & 0 & z & 0 & 0 & -x \\
		0 & 0 & 0 & 0 & 0 & -y \\
		x & 0 & 0 & 0 & 0 & -z \\
		0 & 0 & 0 & 0 & -y^2 & 2xy \\
		0 & 0 & 0 & -y^2 & 0 & 2xy \\
		0 & 0 & x^2 & 0 & 0 & 0 \\
		0 & 0 & 0 & 0 & 0 & -y^2 \\
		z^2 & 0 & 0 & 0 & 0 & 0 \\
	\end{bmatrix}.
\end{equation}
\section{Sensitivity analysis}
\label[appendix]{Ap:sens}

This section presents the sensitivity analysis associated with the proposed formulation. In particular, explicit expressions are derived for the sensitivities of the objective function, transformation matrices, surface normals, and stress field with respect to the design variables. 

\subsection{Sensitivity of enriched-point coordinates}
\label{Ap:SAxn}
The enriched-point coordinates constitute the geometric link between the level set representation and the quantities entering the objective function. In contrast to previous formulations based on a linear interpolation of the level set field \citep{Boom:2021aa, zhang2021on}, the present work employs a quadratic interpolation along the intersected edge. Consequently, the corresponding coordinate sensitivities differ from those reported previously and are derived here. Using the chain rule, these sensitivities are given by
\begin{equation}
	\pdv{\bs{x}_n}{s_j} = \pdv{\bs{x}_n}{\xi_n}\pdv{\xi_n}{\phi_i} \pdv{\phi_i}{s_j}  + \pdv{\bs{x}_n}{\xi_n}\pdv{\xi_n}{\phi_j} \pdv{\phi_j}{s_j} + \pdv{\bs{x}_n}{\xi_n} \pdv{\xi_n}{\phi_k}\pdv{\phi_k}{s_j},
\end{equation}
where $\pdv{\bs{x}_n}{\xi_n} = \frac{1}{2}\left(\bs{p} - \bs{m} \right)$. Note that for linear interpolation of the LSF, the third term doesn't exist. Then, the derivatives of $\xi_n$ with respect to $\phi_i$, $\phi_k$, and $\phi_j$ are obtained analytically by differentiation of \Cref{eq:xiE}, which yields

\begin{equation}
\begin{aligned}
\frac{\partial \xi_n}{\partial \phi_i}
&=
\frac{
(-\phi_k+\phi_j)\sqrt{16\phi_k^2+(-8\phi_i-8\phi_j)\phi_k+(\phi_i-\phi_j)^2}
\pm
\left(
4\phi_k^2
+
(-\phi_i-3\phi_j)\phi_k
-
\phi_j(\phi_i-\phi_j)
\right)
}{
\left(2\phi_k-\phi_j-\phi_i\right)^2 \sqrt{16\phi_k^2+(-8\phi_i-8\phi_j)\phi_k+(\phi_i-\phi_j)^2}
}, \\ 
\frac{\partial \xi_n}{\partial \phi_j}
&=
\frac{
(\phi_k-\phi_i)\sqrt{16\phi_k^2+(-8\phi_i-8\phi_j)\phi_k+(\phi_i-\phi_j)^2}
\pm
\left(
4\phi_k^2
+
(-3\phi_i-\phi_j)\phi_k
+
\phi_i(\phi_i-\phi_j)
\right)
}{
\left(2\phi_k-\phi_j-\phi_i\right)^2 \sqrt{16\phi_k^2+(-8\phi_i-8\phi_j)\phi_k+(\phi_i-\phi_j)^2}
}, \\
\frac{\partial \xi_n}{\partial \phi_k}
&=
-\frac{
(-\phi_i+\phi_j)\sqrt{16\phi_k^2+(-8\phi_i-8\phi_j)\phi_k+(\phi_i-\phi_j)^2}
\pm
\left(
-\phi_i^2
+
4\left(\phi_k-\frac{3\phi_j}{2}\right)\phi_i
+
\phi_j\left(4\phi_k-\phi_j\right)
\right)
}{
\left(2\phi_k-\phi_j-\phi_i\right)^2 \sqrt{16\phi_k^2+(-8\phi_i-8\phi_j)\phi_k+(\phi_i-\phi_j)^2}
}.
\end{aligned}
\end{equation}
Finally, based on \Cref{eq:DesToLSF},
\begin{equation}
	\pdv{\phi_l}{s_j} = \vartheta_j\left(\bs{x}_l\right),
\end{equation}
where $l = m,p,k$.

\subsection{Sensitivity of the objective function with respect to the design variables}
The sensitivity of the objective function with respect to the design variable $s_j$ is given by
\begin{equation}
    \frac{\partial \Phi}{\partial s_j} = \left(\frac{1}{ \abs{\smash{\iota_w}}}\right)^{\frac{1}{p}} \frac{1}{p} \left(\sum_{i \in \iota_w} G_i^p\right)^{\frac{1}{p}-1} \left(\sum_{i \in \iota_w} p \, G_i^{p-1} \frac{\partial G_i}{\partial s_j}\right) ,
    \label{eq:dJdsj}
\end{equation}
where based on \Cref{eq:ERRfinal}
\begin{equation}
\begin{split}
   \frac{\partial G_i}{\partial s_j} = \frac{\pi \epsilon}{2 \mu \bar{E}} &  \biggl( \frac{\partial \bs{Q}_R^\intercal}{\partial s_j} \boldsymbol{\sigma}^\intercal \bs{F}\boldsymbol{\sigma} \bs{Q}_R + \bs{Q}_R^\intercal \frac{\partial \boldsymbol{\sigma}^\intercal}{\partial s_j} \bs{F}\boldsymbol{\sigma} \bs{Q}_R  +  \bs{Q}_R^\intercal \boldsymbol{\sigma}^\intercal \frac{\partial \bs{F}}{\partial s_j}\boldsymbol{\sigma} \bs{Q}_R +  \bs{Q}_R^\intercal \boldsymbol{\sigma}^\intercal \bs{F}\frac{\partial \boldsymbol{\sigma}}{\partial s_j} \bs{Q}_R + \bs{Q}_R^\intercal \boldsymbol{\sigma}^\intercal \bs{F}  \boldsymbol{\sigma} \frac{\partial \bs{Q}_R}{\partial s_j}
    \biggr) .
\end{split}
\end{equation}
This leaves us with the need to compute $\frac{\partial \bs{Q}_R}{\partial s_j}$, $\frac{\partial \bs{F}}{\partial s_j}$, and $\frac{\partial \boldsymbol{\sigma}}{\partial s_j}$. Based on the definitions of $\bs{Q}_R$ and $\bs{F}$ presented in \Cref{Ap:ERR_comp}, the sensitivities of these operators can be obtained through the derivatives of the rotation matrices $\bs{R}$ and $\bs{R}_t$ by means of the chain rule. Therefore, the following derivations focus on the evaluation of $\frac{\partial \bs{R}}{\partial s_j}$, $\frac{\partial \bs{R}_t}{\partial s_j}$, and $\frac{\partial \boldsymbol{\sigma}}{\partial s_j}$.

\subsubsection{Sensitivity of the transformation matrix $\bs{R}$}
\label{Ap:dRds}
From \Cref{eq:R1}, the derivative of $\bs{R}$ with respect to the design variable $s_j$ can be expressed as
\begin{equation}
    \frac{\partial \bs{R}}{\partial s_j}
    =
    \frac{\partial \bs{R}}{\partial n_x} \frac{\partial n_x}{\partial s_j}
    +
    \frac{\partial \bs{R}}{\partial n_y} \frac{\partial n_y}{\partial s_j}
    +
    \frac{\partial \bs{R}}{\partial n_z} \frac{\partial n_z}{\partial s_j},
    \label{eq:dRdn}
\end{equation}
where
{\renewcommand{\arraystretch}{2}
\begin{equation}
	\pdv{\bs{R}}{n_x} =
	\begin{bmatrix}
		\frac{n_x^2-n_y^2-n_z^2}{\left(n_x^2+n_y^2+n_z^2\right)^2} & \frac{2n_x n_y}{\left(n_x^2+n_y^2+n_z^2\right)^2} & \frac{2n_x n_z}{\left(n_x^2+n_y^2+n_z^2\right)^2} \\
		\frac{n_y\left(n_x^4-n_y^2n_z^2-n_z^4\right)}{\left(n_x^2+n_z^2\right)^{\frac{3}{2}}\left(n_x^2+n_y^2+n_z^2\right)^{\frac{3}{2}}} & \frac{n_x n_y^2}{\left(n_x^2+n_z^2\right)^{\frac{1}{2}}\left(n_x^2+n_y^2+n_z^2\right)^{\frac{3}{2}}} & \frac{n_x n_y n_z\left(2n_x^2+n_y^2+2n_z^2\right)}{\left(n_x^2+n_z^2\right)^{\frac{3}{2}}\left(n_x^2+n_y^2+n_z^2\right)^{\frac{3}{2}}} \\
		\frac{n_x n_z}{\left(n_x^2+n_z^2\right)^{\frac{3}{2}}} & 0 & \frac{n_z^2}{\left(n_x^2+n_z^2\right)^{\frac{3}{2}}}
	\end{bmatrix},
\end{equation}
\begin{equation}
	\pdv{\bs{R}}{n_y} =
	\begin{bmatrix}
		\frac{2n_x n_y}{\left(n_x^2+n_y^2+n_z^2\right)^2} & \frac{-n_x^2+n_y^2-n_z^2}{\left(n_x^2+n_y^2+n_z^2\right)^2} & \frac{2n_y n_z}{\left(n_x^2+n_y^2+n_z^2\right)^2} \\
		-\frac{n_x\sqrt{n_x^2+ n_z^2}}{\left(n_x^2+n_y^2+n_z^2\right)^{\frac{3}{2}}} & -\frac{n_y\sqrt{n_x^2+ n_z^2}}{\left(n_x^2+n_y^2+n_z^2\right)^{\frac{3}{2}}} & -\frac{n_z\sqrt{n_x^2+ n_z^2}}{\left(n_x^2+n_y^2+n_z^2\right)^{\frac{3}{2}}} \\
		0 & 0 & 0
	\end{bmatrix},
\end{equation}
\begin{equation}
	\pdv{\bs{R}}{n_z} =
	\begin{bmatrix}
		\frac{2n_x n_z}{\left(n_x^2+n_y^2+n_z^2\right)^2} & \frac{2n_y n_z}{\left(n_x^2+n_y^2+n_z^2\right)^2} & \frac{-n_x^2-n_y^2+n_z^2}{\left(n_x^2+n_y^2+n_z^2\right)^2} \\
		\frac{n_x n_y n_z\left(2n_x^2+n_y^2+2n_z^2\right)}{\left(n_x^2+n_z^2\right)^{\frac{3}{2}}\left(n_x^2+n_y^2+n_z^2\right)^{\frac{3}{2}}} & \frac{n_y^2 n_z}{\sqrt{n_x^2+n_z^2}\left(n_x^2+n_y^2+n_z^2\right)^{\frac{3}{2}}} & -\frac{n_y\left(n_x^4+n_y^2n_x^2-n_z^4\right)}{\left(n_x^2+n_z^2\right)^{\frac{3}{2}}\left(n_x^2+n_y^2+n_z^2\right)^{\frac{3}{2}}} \\
		-\frac{n_x^2}{\left(n_x^2+n_z^2\right)^\frac{3}{2}} & 0 & -\frac{n_x n_z}{\left(n_x^2+n_z^2\right)^\frac{3}{2}}.
	\end{bmatrix}.
\end{equation}
}
The derivatives of the normal components with respect to the design variables are given by
\begin{equation}
    \begin{bmatrix}
        \frac{\partial n_x}{\partial s_j} \\
        \frac{\partial n_y}{\partial s_j} \\
        \frac{\partial n_z}{\partial s_j}
    \end{bmatrix}
    =
    \begin{bmatrix}
        \sum \frac{\partial n_x}{\partial n_{ix}} \frac{\partial n_{ix}}{\partial s_j} + \sum \frac{\partial n_x}{\partial n_{iy}} \frac{\partial n_{iy}}{\partial s_j} + \sum \frac{\partial n_x}{\partial n_{iz}} \frac{\partial n_{iz}}{\partial s_j} \\
        \sum \frac{\partial n_y}{\partial n_{ix}} \frac{\partial n_{ix}}{\partial s_j} + \sum \frac{\partial n_y}{\partial n_{iy}} \frac{\partial n_{iy}}{\partial s_j} + \sum \frac{\partial n_y}{\partial n_{iz}} \frac{\partial n_{iz}}{\partial s_j} \\
        \sum \frac{\partial n_z}{\partial n_{ix}} \frac{\partial n_{ix}}{\partial s_j} + \sum \frac{\partial n_z}{\partial n_{iy}} \frac{\partial n_{iy}}{\partial s_j} + \sum \frac{\partial n_z}{\partial n_{iz}} \frac{\partial n_{iz}}{\partial s_j} \\
    \end{bmatrix},
\end{equation}
where
\begin{equation}
    \frac{\partial n_x}{\partial n_{ix}} = \frac{\left(\sum n_{iy}\right)^2 + \left(\sum n_{iz}\right)^2}{N_s [\left(\sum n_{ix}\right)^2 + \left(\sum n_{iy}\right)^2 + \left(\sum n_{iz}\right)^2]^{\frac{3}{2}}},
\end{equation}

\begin{equation}
    \frac{\partial n_y}{\partial n_{iy}} = \frac{\left(\sum n_{ix}\right)^2 + \left(\sum n_{iz}\right)^2}{N_s [\left(\sum n_{ix}\right)^2 + \left(\sum n_{iy}\right)^2 + \left(\sum n_{iz}\right)^2]^{\frac{3}{2}}},
\end{equation}

\begin{equation}
    \frac{\partial n_z}{\partial n_{iz}} = \frac{\left(\sum n_{ix}\right)^2 + \left(\sum n_{iy}\right)^2}{N_s [\left(\sum n_{ix}\right)^2 + \left(\sum n_{iy}\right)^2 + \left(\sum n_{iz}\right)^2]^{\frac{3}{2}}},
\end{equation}

\begin{equation}
    \frac{\partial n_x}{\partial n_{iy}} = \frac{\partial n_y}{\partial n_{ix}} = - \frac{\sum n_{ix} \sum n_{iy}}{N_s [\left(\sum n_{ix}\right)^2 + \left(\sum n_{iy}\right)^2 + \left(\sum n_{iz}\right)^2]^{\frac{3}{2}}},
\end{equation}

\begin{equation}
    \frac{\partial n_x}{\partial n_{iz}} = \frac{\partial n_z}{\partial n_{ix}} = - \frac{\sum n_{ix} \sum n_{iz}}{N_s [\left(\sum n_{ix}\right)^2 + \left(\sum n_{iy}\right)^2 + \left(\sum n_{iz}\right)^2]^{\frac{3}{2}}},
\end{equation}

\begin{equation}
    \frac{\partial n_y}{\partial n_{iz}} = \frac{\partial n_z}{\partial n_{iy}} =  - \frac{\sum n_{iy} \sum n_{iz}}{N_s [\left(\sum n_{ix}\right)^2 + \left(\sum n_{iy}\right)^2 + \left(\sum n_{iz}\right)^2]^{\frac{3}{2}}}.
\end{equation}
To evaluate $\frac{\partial \bs{n}_i}{\partial s_j}$, let
$\bs{D}=\bs{d}_{01}\times\bs{d}_{02}$, such that
\Cref{eq:nd} can be rewritten as
\begin{equation}
    \bs{n}_i
    =
    m_i \bs{D}
    =
    m_i \left(D_x,D_y,D_z\right).
\end{equation}
Consequently,
\begin{equation}
    \frac{\partial \bs{n}_i}{\partial s_j}
    =
    m_i\frac{\partial \bs{D}}{\partial s_j},
\end{equation}
or, componentwise,
\begin{equation}
    \begin{bmatrix}
        \dfrac{\partial n_{ix}}{\partial s_j}\\[2mm]
        \dfrac{\partial n_{iy}}{\partial s_j}\\[2mm]
        \dfrac{\partial n_{iz}}{\partial s_j}
    \end{bmatrix}
    =
    m_i
    \begin{bmatrix}
        \dfrac{\partial D_x}{\partial s_j}\\[2mm]
        \dfrac{\partial D_y}{\partial s_j}\\[2mm]
        \dfrac{\partial D_z}{\partial s_j}
    \end{bmatrix}.
\end{equation}
The derivative of $\bs D$ follows from the product rule for the
cross product:
\begin{equation}
\begin{aligned}
    \frac{\partial \bs{D}}{\partial s_j}
    &=
    \frac{\partial \bs{d}_{01}}{\partial s_j}
    \times \bs{d}_{02}
    +
    \bs{d}_{01}
    \times
    \frac{\partial \bs{d}_{02}}{\partial s_j},\\
    \frac{\partial \bs{d}_{01}}{\partial s_j}
    &=
    \frac{\partial \bs{x}_1}{\partial s_j}
    -
    \frac{\partial \bs{x}_0}{\partial s_j},\\
    \frac{\partial \bs{d}_{02}}{\partial s_j}
    &=
    \frac{\partial \bs{x}_2}{\partial s_j}
    -
    \frac{\partial \bs{x}_0}{\partial s_j}.
\end{aligned}
\end{equation}
Here, $\frac{\partial \bs{x}_i}{\partial s_j}$ is evaluated as
described in \Cref{Ap:SAxn}.

\subsubsection{Sensitivity of the transformation matrix $\bs{R}_t$}
\label{Ap:dRTds}
Since $\bs{R}_t$ depends on the principal stress direction $\theta_p$, its sensitivity with respect to the design variables can be obtained through the chain rule as
\begin{equation}
    \begin{split}
        \pdv{\bs{R}_t}{s_j} &= \frac{\partial \bs{R}_t}{\partial \theta_p} \left( \pdv{\theta_p}{\sigma_{y'z'}} \pdv{\sigma_{y'z'}}{s_j} + \pdv{\theta_p}{\sigma_{y'y'}} \pdv{\sigma_{y'y'}}{s_j} + \pdv{\theta_p}{\sigma_{z'z'}} \pdv{\sigma_{z'z'}}{s_j} \right) \\
        &= \frac{\partial \bs{R}_t}{\partial \theta_p} \left(
        \begin{bmatrix}
            \pdv{\theta_p}{\sigma_{y'z'}} & \pdv{\theta_p}{\sigma_{y'y'}} & \pdv{\theta_p}{\sigma_{z'z'}}
        \end{bmatrix}
        \begin{bmatrix}
            \pdv{\sigma_{y'z'}}{s_j} \\
            \pdv{\sigma_{y'y'}}{s_j} \\
            \pdv{\sigma_{z'z'}}{s_j}
        \end{bmatrix}\right).
    \end{split}
    \label{eq:dRtdsj}
\end{equation}

In the above equation,
\begin{equation}
    \pdv{\bs{R}_t}{\theta_p}
    =
    \begin{bmatrix}
        0 & 0 & 0 \\
        0 & -\sin\left(\theta_p\right) & -\cos\left(\theta_p\right) \\
        0 & \cos\left(\theta_p\right) & -\sin\left(\theta_p\right)
    \end{bmatrix},
\end{equation}
\begin{equation}
    \pdv{\theta_p}{\sigma_{y'z'}} =  \frac{\sigma_{y'y'}-\sigma_{z'z'}}{\left(\sigma_{z'z'}-\sigma_{y'y'}\right)^2 + 4\sigma_{y'z'}^2},
\end{equation}
\begin{equation}
    \pdv{\theta_p}{\sigma_{y'y'}} =  -\frac{\sigma_{y'z'}}{\left(\sigma_{z'z'}-\sigma_{y'y'}\right)^2 + 4\sigma_{y'z'}^2},
\end{equation}
\begin{equation}
    \pdv{\theta_p}{\sigma_{z'z'}} =  \frac{\sigma_{y'z'}}{\left(\sigma_{z'z'}-\sigma_{y'y'}\right)^2 + 4\sigma_{y'z'}^2}.
\end{equation}
The local stress components entering \Cref{eq:dRtdsj} are obtained from the transformed stress tensor
\begin{equation}
    \boldsymbol{\sigma}'=\bs{R}\boldsymbol{\sigma}\bs{R}^{\intercal}.
\end{equation}
Accordingly,
\begin{equation}
    \sigma_{y'z'}=\sigma'_{23},
    \qquad
    \sigma_{y'y'}=\sigma'_{22},
    \qquad
    \sigma_{z'z'}=\sigma'_{33}.
\end{equation}
Differentiating the transformed stress tensor with respect to the design variable yields
\begin{equation}
    \frac{\partial \boldsymbol{\sigma}'}{\partial s_j}
    =
    \frac{\partial \bs{R}}{\partial s_j}
    \boldsymbol{\sigma}
    \bs{R}^{\intercal}
    +
    \bs{R}
    \frac{\partial \boldsymbol{\sigma}}{\partial s_j}
    \bs{R}^{\intercal}
    +
    \bs{R}
    \boldsymbol{\sigma}
    \frac{\partial \bs{R}^{\intercal}}{\partial s_j}.
\end{equation}
Here, $\frac{\partial \bs{R}}{\partial s_j}$ is available from \Cref{Ap:dRds}, whereas the sensitivity of the recovered nodal stress field, $\frac{\partial \boldsymbol{\sigma}}{\partial s_j}$, is derived in \Cref{Ap:SA_stress}.

\subsubsection{Sensitivity of the recovered nodal stress}
\label{Ap:SA_stress}
The sensitivity of the recovered nodal stress with respect to the design variable $s_j$ is expressed as

\begin{equation}
    \frac{\partial \bs{\sigma}\left(\bs{x}_i\right)}{\partial s_j} = \frac{\partial \bs{\sigma}\left(\bs{x}_i\right)}{\partial \bs{x}_n} \frac{\partial \bs{x}_n}{\partial s_j} = \frac{1}{\left| \mathcal{P}_i \right|} \frac{\partial \left(\sum \bs{E}_{\sigma}\left(\bs{x}_i\right) \hat{\bs{\sigma}}_e\right)}{\partial \bs{x}_n} \frac{\partial \bs{x}_n}{\partial s_j} = \frac{1}{\left| \mathcal{P}_i \right|} \sum \left(\frac{\partial \bs{E}_{\sigma}\left(\bs{x}_i\right)}{\partial \bs{x}_n} \hat{\bs{\sigma}}_e + \bs{E}_{\sigma}\left(\bs{x}_i\right)\frac{\partial \hat{\bs{\sigma}}_e}{\partial \bs{x}_n}\right) \frac{\partial \bs{x}_n}{\partial s_j}.
\end{equation}
The derivatives of $\bs{E}_{\sigma}$ with respect to $\bs{x}_n$ are given by

\begin{equation}
	\frac{\partial \bs{E}_{\sigma}}{\partial \bs{x}_n} = 
	\begin{bmatrix}
		\pdv{\bs{e}_{\sigma}}{\bs{x}_n} & & & & & \\
		& \pdv{\bs{e}_{\sigma}}{\bs{x}_n} & & & & \\
		& & \pdv{\bs{e}_{\sigma}}{\bs{x}_n} & & & \\
		& & & \pdv{\bs{e}_{\sigma}}{\bs{x}_n} & & \\
		& & & & \pdv{\bs{e}_{\sigma}}{\bs{x}_n} & \\
		& & & & & \pdv{\bs{e}_{\sigma}}{\bs{x}_n}
	\end{bmatrix},
\end{equation}
where $\frac{\partial \bs{E}_{\sigma}}{\partial \bs{x}_n}$ denotes $\pdv{\bs{E}_{\sigma}}{x_n}$, $\pdv{\bs{E}_{\sigma}}{y_n}$ and $\pdv{\bs{E}_{\sigma}}{z_n}$, while $\pdv{\bs{e}_{\sigma}}{\bs{x}_n}$ denotes $\pdv{\bs{e}_{\sigma}}{x_n}$, $\pdv{\bs{e}_{\sigma}}{y_n}$ and $\pdv{\bs{e}_{\sigma}}{z_n}$. The derivatives of the polynomial basis vectors $\bs{e}_{\sigma}$ and consequently those of the interpolation matrices $\bs{E}_{\sigma}$ and $\bs{E}_{\zeta}$, are obtained by straightforward differentiation and are omitted for brevity.
To evaluate $\frac{\partial \hat{\bs{\sigma}}_e}{\partial \bs{x}_n}$, let
\begin{equation}
	\bs{A}_e = \begin{bmatrix}
		\int_e \bar{\bs{E}}_{\sigma}^\intercal \bs{E}_{\sigma} \, \dd{e} \\
		\int_e \bs{E}_{\zeta}^\intercal \bs{\partial}_{\sigma} \bs{E}_{\sigma} \, \dd{e}
	\end{bmatrix}, \; \bs{B}_{e1} = \int_e \bar{\bs{E}}_{\sigma}^\intercal \bs{\sigma}_e^h \, \dd{e}, \; \bs{B}_{e2} = -\int_e \bs{E}_{\zeta}^\intercal \bs{b} \, \dd{e}.
    \label{eq:AB}
\end{equation}
Accordingly, the derivative of \Cref{eq:stress} can be expressed as
\begin{equation}
	\left(\sum_{e\in \varepsilon_i} \pdv{\bs{A}_e}{\bs{x}_n}\right)\hat{\bs{\sigma}}_e + \left(\sum_{e\in \varepsilon_i}\bs{A}_e\right) \pdv{\hat{\bs{\sigma}}_e}{\bs{x}_n} = \sum_{e\in \varepsilon_i} \begin{bmatrix}
		\pdv{\bs{B}_{e1}}{\bs{x}_n} \\
		\pdv{\bs{B}_{e2}}{\bs{x}_n}
	\end{bmatrix}.
    \label{eq:dstress}
\end{equation}
Using \Cref{eq:dstress}, the derivative of $\hat{\bs{\sigma}}_e$ with respect to $\bs{x}_n$ is obtained as
\begin{equation}
	\pdv{\hat{\bs{\sigma}}_e}{\bs{x}_n} = \left(\sum_{e\in \varepsilon_i}\bs{A}_e\right)^{-1} \left[\sum_{e\in \varepsilon_i} \begin{bmatrix}
		\pdv{\bs{B}_{e1}}{\bs{x}_n} \\
		\pdv{\bs{B}_{e2}}{\bs{x}_n}
	\end{bmatrix} - \left(\sum_{e\in \varepsilon_i} \pdv{\bs{A}_e}{\bs{x}_n}\right)\hat{\bs{\sigma}}_e\right],
\end{equation}
where the derivatives of $\bs{A}_e$, $\bs{B}_{e1} $ and $\bs{B}_{e2}$ are obtained by differentiating the corresponding integrals using the derivatives of $\bs{E}_{\zeta}$ and $\bs{E}_{\sigma}$.

\subsection{Sensitivity of the objective function with respect to the state variables}
The sensitivity of the objective with respect to the displacement field is given by
\begin{equation}
    \frac{\partial \Phi}{\partial \bs{U}} = \left(\frac{1}{ \abs{\iota_w}}\right)^{\frac{1}{p}} \frac{1}{p} \left(\sum_{i \in \iota_w} G_i^p\right)^{\frac{1}{p}-1} \left(\sum_{i \in \iota_w} p G_i^{p-1} \frac{\partial G_i}{\partial \bs{U}}\right) ,
    \label{eq:dJduj}
\end{equation}
where,
\begin{equation}
\begin{split}
   \frac{\partial G_i}{\partial \bs{U}} = \frac{\pi \epsilon}{2 \mu \bar{E}} &  \biggl( \frac{\partial \bs{Q}_R^\intercal}{\partial \bs{U}} \boldsymbol{\sigma}^\intercal \bs{F}\boldsymbol{\sigma} \bs{Q}_R + \bs{Q}_R^\intercal \frac{\partial \boldsymbol{\sigma}^\intercal}{\partial \bs{U}} \bs{F}\boldsymbol{\sigma} \bs{Q}_R  +  \bs{Q}_R^\intercal \boldsymbol{\sigma}^\intercal \frac{\partial \bs{F}}{\partial \bs{U}}\boldsymbol{\sigma} \bs{Q}_R +  \bs{Q}_R^\intercal \boldsymbol{\sigma}^\intercal \bs{F}\frac{\partial \boldsymbol{\sigma}}{\partial \bs{U}} \bs{Q}_R + \bs{Q}_R^\intercal \boldsymbol{\sigma}^\intercal \bs{F}  \boldsymbol{\sigma} \frac{\partial \bs{Q}_R}{\partial \bs{U}}
    \biggr) .
\end{split}
\end{equation}
The evaluation of $\frac{\partial G_i}{\partial \bs{U}}$ requires the sensitivities $\frac{\partial \bs{Q}_R}{\partial U}$, $\frac{\partial \bs{F}}{\partial U}$, and
$\frac{\partial \boldsymbol{\sigma}}{\partial U}$. Based on the definitions of $\bs{Q}_R$ and $\bs{F}$, the sensitivities of these operators can be obtained through the derivative of the transformation matrix $\bs{R}_t$. Unlike the design sensitivity analysis, the transformation matrix $\bs{R}$ is independent of the state variables and therefore does not contribute to the present derivation. Consequently, the following expressions focus on the evaluation of
$\frac{\partial \bs{R}_t}{\partial \bs{U}}$ and $\frac{\partial \boldsymbol{\sigma}}{\partial \bs{U}}$. The sensitivity of the rotation matrix $\bs{R}_t$ is given by

\begin{equation}
    \begin{split}
        \pdv{\bs{R}_t}{U} = \frac{\partial \bs{R}_t}{\partial \theta_p} \left(
        \begin{bmatrix}
            \pdv{\theta_p}{\sigma_{y'z'}} & \pdv{\theta_p}{\sigma_{y'y'}} & \pdv{\theta_p}{\sigma_{z'z'}}
        \end{bmatrix}
        \begin{bmatrix}
            \pdv{\sigma_{y'z'}}{\bs{U}} \\
            \pdv{\sigma_{y'y'}}{\bs{U}} \\
            \pdv{\sigma_{z'z'}}{\bs{U}}
        \end{bmatrix}\right),
    \end{split}
    \label{eq:dRtdU}
\end{equation}
where $\frac{\partial \bs{R}_t}{\partial \theta_p}$ and $\frac{\partial \theta_p}{\partial \boldsymbol{\sigma}}$ were derived in \Cref{Ap:dRTds}. It therefore remains to evaluate $\frac{\partial \boldsymbol{\sigma}}{\partial \bs{U}}$. Since $\bs{E}_{\sigma}(\bs{x}_i)$ is independent of the displacement vector $\bs{U}$, the recovered nodal stress sensitivity is given by
\begin{equation}
    \frac{\partial \bs{\sigma}(\bs{x}_i)}{\partial \bs{U}}
    =
    \frac{1}{\left|\mathcal{P}_i\right|}
    \frac{\partial}{\partial \bs{U}}
    \left(
        \sum_{e\in\mathcal{P}_i}
        \bs{E}_{\sigma}(\bs{x}_i)\hat{\bs{\sigma}}_e
    \right)
    =
    \frac{1}{\left|\mathcal{P}_i\right|}
    \sum_{e\in\mathcal{P}_i}
    \bs{E}_{\sigma}(\bs{x}_i)
    \frac{\partial\hat{\bs{\sigma}}_e}{\partial\bs{U}},
\end{equation}
where
\begin{equation}
	\pdv{\hat{\bs{\sigma}}_e}{\bs{U}} = \left(\sum_{e\in \varepsilon_i}\bs{A}_e\right)^{-1} \left[\sum_{e\in \varepsilon_i} \begin{bmatrix}
		\pdv{\bs{B}_{e1}}{\bs{U}} \\
		\pdv{\bs{B}_{e2}}{\bs{U}}
	\end{bmatrix} - \left(\sum_{e\in \varepsilon_i} \pdv{\bs{A}_e}{\bs{U}}\right)\hat{\bs{\sigma}}_e\right]
    =
    \left(\sum_{e\in \varepsilon_i}\bs{A}_e\right)^{-1} \sum_{e\in \varepsilon_i} \begin{bmatrix}
    	\pdv{\bs{B}_{e1}}{\bs{U}} \\
    	\pdv{\bs{B}_{e2}}{\bs{U}}
    \end{bmatrix}.
\end{equation}
The second term vanishes because, according to \Cref{eq:AB}, $\bs{A}_e$ is independent of the displacement vector $\bs{U}$. The remaining derivatives of $\bs{B}_{e1}$ and $\bs{B}_{e2}$ follow directly from their definitions.

\subsection{The sensitivity of the stiffness matrix with respect to the design variables}
According to \Cref{eq:ke}, the derivative of element stiffness $\bs{k}_e$ with respect to $\bs{x}_n$ is
\begin{equation}
	\pdv{\bs{k}_e}{\bs{x}_n} = \int_e \left(\pdv{j_e}{\bs{x}_n}\bs{B}^\intercal \bs{D} \bs{B} + j_e\pdv{\bs{B}^\intercal}{\bs{x}_n}\bs{D}\bs{B} + j_e\bs{B}^\intercal\bs{D}\pdv{\bs{B}}{\bs{x}_n}\right) \, \dd{\bs{\xi}},
	\label{eq:dkdx_e}
\end{equation}
where $\bs{B}$ is defined in \Cref{sec:elas}. Its derivative is given by
\begin{equation}
\pdv{\bs{B}}{\bs{x}_n}
=
\begin{bmatrix}
\bs{0} &
\bs{0} &
\cdots &
\bs{0} &
\pdv{\Delta \bs{\psi}_l}{\bs{x}_n}
&
\cdots &
\pdv{\Delta \bs{\psi}_m}{\bs{x}_n}
\end{bmatrix}.
\label{eq:dBdx}
\end{equation}
Note that only the enriched part of the formulation contributes to the sensitivity, since the derivatives of the background shape functions of the linear tetrahedral element are constant throughout the integration element and therefore do not depend on the enriched node position. The derivative of the enrichment function is calculated as
\begin{equation}
\label{eq:denrich}
\pdv{}{\bs{x}_n}\!\left(\nabla_{\bs{x}}{\psi}_i\right)
=
\pdv{\bs{J}_e^{-1}}{\bs{x}_n}
\nabla_{\bs{\xi}}{\psi}_i
+
\cancel{\bs{J}_e^{-1}
\pdv{\nabla_{\bs{\xi}}{\psi}_i}{\bs{x}_n}}
\end{equation}
The second term in \Cref{eq:denrich} vanishes because the enrichment function is evaluated at an integration point defined in the parent coordinate system. Since the location of the integration point is fixed in the parent coordinates, the value of the enrichment function remains unchanged regardless of the position of the integration element in the physical domain. Therefore, the sensitivity of the enrichment gradient reduces to the sensitivity of the inverse Jacobian, which is obtained using the matrix inverse differentiation rule,
\begin{equation}
\label{eq:derJe}
    \pdv{ \bs{J}_e^{-1}}{\bs{x}_n} = -\bs{J}_e^{-1} \pdv{\bs{J}_e}{\bs{x}_n} \bs{J}_e^{-1},
\end{equation}
where $\pdv{\bs{J}_e}{\bs{x}_n}$ can be found in \Cref{Ap:normalJ}.

\section{Isoparametric mapping of integration elements and normal Jacobian}
\label[appendix]{Ap:normalJ}

The sensitivity analysis further requires the derivative of the Jacobian determinant. Using Jacobi's formula \citep{magnus2019matrix}, the derivative of a determinant is obtained as the trace of the product of the adjugate matrix and the matrix derivative. Noting that $\left(\mathrm{adj}\left(\bs{J}_e\right)=j_e \bs{J}_e^{-T}\right)$, the derivative of the element Jacobian determinant becomes 
\begin{equation}
\label{eq:derdet}
	\pdv{j_e}{\bs{x}_n} = \mathrm{Tr}\left(\mathrm{adj}\left(\bs{J}_e\right) \pdv{\bs{J}_e}{\bs{x}_e}\right)
\end{equation}

For both \Cref{eq:derJe} and \Cref{eq:derdet} the sensitivity of the Jacobian is required. Since the element Jacobian is given by $\bs{J}_e = \bs{x}^\intercal_e \pdv{\bs{N}_e}{\bs{\xi}_e}$, its derivative with respect to the enriched node coordinates can be expressed as
\begin{equation}
	\pdv{\bs{J}_e}{\bs{x}_n} = \pdv{\bs{x}_e^\intercal}{\bs{x}_n} \pdv{\bs{N}_e}{\bs{\xi}_e} + \cancel{\bs{x}_e^\intercal\frac{\partial^2\bs{N}_e}{\partial \bs{\xi}_e \partial \bs{x}_n}},
\end{equation}
where $\pdv{\bs{x}_e^\intercal}{\bs{x}_n}$ is a selection matrix containing zeros everywhere except for unit entries corresponding to the degrees of freedom of enriched node $n$.

\section{Sensitivity verification}
\label[appendix]{Ap:sensVer}

The sensitivity of the topology optimization formulation (\Cref{Ap:sens}) is verified using finite difference analysis. The analytical sensitivity, $\pdv{\Lambda}{s_j}$, is compared to its finite difference approximation, $J'_j$. The relative error is defined as

\begin{equation}
\delta_j = \frac{J'_j - \pdv{\Lambda}{s_j}}{J'_j} .
\end{equation}

\begin{figure}[t]
  \centering
  \begin{subfigure}{0.48\textwidth}
    \centering
    \includegraphics[width=\linewidth]{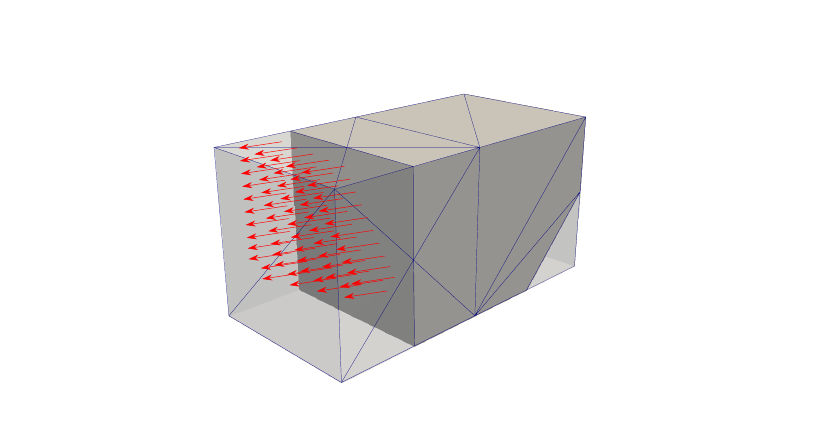}
    \caption{}
    \label{fig:FDA_a}
  \end{subfigure}
  \begin{subfigure}{0.48\textwidth}
    \centering
    \includegraphics[width=\linewidth]{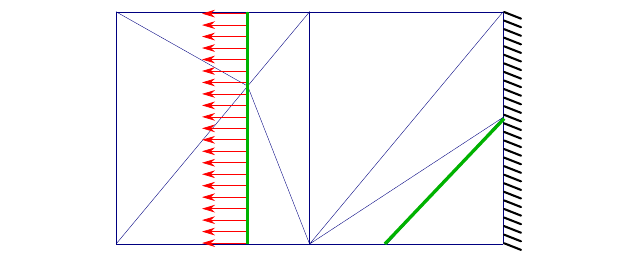}
    \caption{}
    \label{fig:FDA_b}
  \end{subfigure} \\
   \begin{subfigure}{0.32\textwidth}
    \centering
    \makebox[\linewidth][c]{%
    \hspace*{-0.8cm}\includegraphics{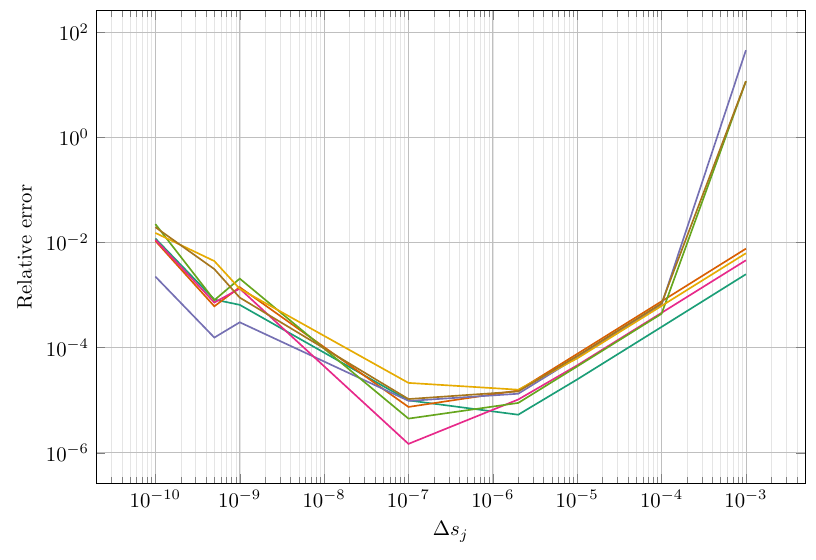}%
}
    \caption{}
    \label{fig:FDA_c}
  \end{subfigure}
  \caption{Sensitivity analysis verification problem. (a) Three-dimensional view of the domain, where the transparent and opaque regions denote the void and solid phases, respectively. A uniformly distributed traction is applied on the internal face and the structure is fixed on the opposite boundary; (b) Corresponding side view; the green line indicates the LSF cut; (c) Relative error between analytical and finite-difference sensitivities as a function of the perturbation size $\Delta s_j$.}
  \label{fig:FDA_start}
\end{figure}

\Cref{fig:FDA_a}–\Cref{fig:FDA_c} illustrate the geometry employed for sensitivity verification together with the corresponding results. A three-dimensional view of the structure is presented in \Cref{fig:FDA_a}, where transparency is used to visualize the immersed configuration. For clarity, a two-dimensional representation is provided in \Cref{fig:FDA_b}, highlighting that the loads are applied on an enriched surface, while the green lines denote the interface defined by the LSF. In addition to the primary vertical loading surface, an additional inclined cut is introduced to assess the SA on a non-aligned plane, as shown in the bottom-right corner. The Young’s moduli of the solid and void phases are taken as $E_s = 1$ and $E_v = 10^{-6}$, respectively, and a distributed load of magnitude $F = 1$ is applied. The relative differences as a function of the finite difference step size $\Delta s_j$ are reported in \Cref{fig:FDA_c}. The close agreement between analytical and numerical sensitivities demonstrates the accuracy of the proposed SA framework.

\section*{Acknowledgments}
The authors gratefully acknowledge financial support from the Dutch Research Council (NWO), Nederlandse Organisatie voor Wetenschappelijk Onderzoek, under award number KICH1.ST04.22.005.

\bibliographystyle{abbrvnat}
\bibliography{references}

\end{document}